\documentclass{emulateapj}
\usepackage{apjfonts}
\usepackage{bm}
\usepackage{amsmath}
\usepackage{rotating}

\newcommand\lsim{\mathrel{\rlap{\lower4pt\hbox{\hskip1pt$\sim$}}
        \raise1pt\hbox{$<$}}}
\newcommand\gsim{\mathrel{\rlap{\lower4pt\hbox{\hskip1pt$\sim$}}
        \raise1pt\hbox{$>$}}}
\newcommand\propsim{\mathrel{\rlap{\lower4pt\hbox{\hskip1pt$\sim$}}
        \raise1pt\hbox{$\propto$}}}

\newcommand{\D}{\mathrm{d}}

\newcommand{\fast}{\mathrm{fast}}
\newcommand{\slow}{\mathrm{slow}}
\newcommand{\spin}{\mathrm{spin}}

\newcommand{\dL}{d_{\rm L}}
\newcommand{\yr}{\,\mathrm{yr}}
\newcommand{\dy}{\,\mathrm{days}}

\newcommand{\Mpc}{\,\mathrm{Mpc}}

\newcommand{\mHz}{\,\mathrm{mHz}}
\newcommand{\Hz}{\,\mathrm{Hz}}
\newcommand{\Mchirp}{\mathcal{M}}
\newcommand{\Msun}{\mathrm{M}_{\odot}}

\newcommand{\ti}{t_{\mathrm{i}}}
\newcommand{\tf}{t_{\mathrm{f}}}

\begin{document}

\title{Pre-Merger Localization of Gravitational-Wave Standard Sirens
With {\it LISA}:\\Triggered Search for an Electromagnetic Counterpart}

\author{Bence Kocsis}
\affiliation{Harvard-Smithsonian Center for Astrophysics, 60 Garden Street, Cambridge, MA 02138}
\email{bkocsis@cfa.harvard.edu}
\author{Zolt\'an Haiman}
\affiliation{Department of Astronomy, Columbia University, 550 West 120th Street, New York, NY 10027}
\email{zoltan@astro.columbia.edu}
\author{Kristen Menou}
\affiliation{Department of Astronomy, Columbia University, 550 West 120th Street, New York, NY 10027}
\email{kristen@astro.columbia.edu}


\begin{abstract}
  Electromagnetic (EM) counterparts to supermassive black hole binary mergers observed by {\it LISA} can be localized to within the field of view of astronomical instruments ($\sim 10\deg^2$) hours to weeks prior to coalescence. The temporal coincidence of any prompt EM counterpart with a gravitationally-timed merger may offer the best chance of identifying a unique host galaxy. We discuss the challenges posed by searches for such prompt EM counterparts and propose novel observational strategies to address them.  In particular, we discuss the size and shape evolution of the LISA localization error ellipses on the sky, and quantify the corresponding requirements for dedicated EM surveys of the area prior to coalescence.  The basic requirements of a wide field of view and fast detectors are similar to those for searches being planned for distant cosmological supernovae.  A triggered EM counterpart search campaign will require monitoring a several--square degree area. It could aim for variability at the 24--27 mag level in optical bands, for example, which corresponds to 1-10\% of the Eddington luminosity of the prime {\it LISA} sources of $\sim (10^6$--$10^7)\,{\rm M_\odot}$ BHs at $z=1-2$, on time--scales of minutes to hours, the orbital time--scale of the binary in the last 2--4 weeks of coalescence. A cross--correlation of the period of any variable EM signal with the quasi--periodic gravitational waveform over 10-1000 cycles may aid the detection. Alternatively, EM searches can detect a transient signal accompanying the coalescence.  The triggered searches will be ambitious, but if they successfully identify a unique prompt electromagnetic counterpart, they will enable new fundamental tests of gravitational physics.  We highlight the measurement of differences in the arrival times of photons and gravitons from the same cosmological source as a valuable independent test of the massive character of gravity, and of possible violations of Lorentz invariance in the gravity sector. 
\end{abstract}

\section{Introduction}

The detection by the future {\it Laser Interferometric Space Antenna}
({\it LISA}) of gravitational waves (GW) emitted during the
coalescence of supermassive black holes (SMBHs) in the mass range
$\sim (10^4$--$10^7)\,{\rm M_\odot}/(1+z)$ will constitute a milestone
for fundamental physics and astrophysics.  High angular resolution
observations of the nuclei of several dozen galaxies over the past
several years have revealed that supermassive black holes (SMBHs) are
ubiquitous in local galaxies (see, e.g., the review by
\citealt{kr95}).  Mergers are expected to play an important role in
the formation and evolution of SMBHs on cosmological time--scales.
The number of events that will be detectable by {\it LISA} is
uncertain by orders of magnitude, depending on model assumptions
\citep{haehnelt94,mhn01,vhm03,mic07,svh07,its04}.  However, for many
events in the above mass and redshift range, {\it LISA} would be able
to measure the masses and spin vectors of the SMBHs, their orbital
parameters, and their luminosity distance, to a precision
unprecedented in any other type of astronomical observation
\citep[e.g.][]{vec04,lh06}.  In addition to providing crucial new
information for hierarchical structure formation scenarios and black
hole astrophysics, the observation of these GW sources would offer new
tests of general relativity and cosmology
\citep{hm05,bbw05,bbw05b,Arun06a,Arun06b}.

While the GW signatures themselves are a rich source of information,
the secure identification of the LISA source in electromagnetic (EM)
wavebands would enable a host of new applications. Traditional
astronomical tools could be used to study the nature of these sources
in detail, clarifying how the SMBHs grow, evolve, and affect their
galaxy environments \citep[e.g.][]{koc06}.  The simultaneous study of
photons and gravitons from a single source would also likely probe
fundamental aspects of gravitational physics \citep[e.g.][]{dm07}.

The single most important unresolved issue is whether counterparts to
{\it LISA} sources will, in fact, be found in electromagnetic (EM)
wavebands.  The final angular localization by {\it LISA} itself is
relatively poor (typically a few $\times$ 0.1 square degrees at
merger), and early studies on counterparts pointed out that this large
solid angle will contain many thousands of galaxies, suggesting that
identifying the galactic host of the coalescing SMBHs will be
difficult, or impossible \citep[e.g.][]{vec04}.  However, recent work
has reached much more optimistic conclusions.  Using {\it LISA}'s
localization of the source within a 3-D error volume (rather than 2-d
sky position) reduces the number of candidate galaxies by $\sim 2-3$
orders of magnitude \citep{hh05,koc06}.  Furthermore, if the
coalescing SMBHs themselves produce bright emission (such as bright
quasars), or are associated with some other, similarly rare subset of
all galaxies (such as ultra--luminous infrared galaxies [ULIRGs],
whose large infrared luminosity may arise from the galaxy merger
process that also causes the SMBH binary coalescence), then the
identification of a unique counterpart may become feasible for a
typical {\it LISA} source \citep{koc06}.

Another possibility, raised by \citet[hereafter Paper I]{paper1}, is to
monitor the sky for EM counterparts in real time, even as the SMBH
inspiral proceeds. Various strategies to carry out this monitoring
effort are the subject of the present follow--up paper.  If the exact
nature of counterparts or host galaxies were securely known from
ab--initio models, then one could hope to identify them post-merger,
e.g. based on the peculiar photometric or spectroscopic signatures.
However, such ab--initio predictions are difficult. Searching for a
variable or transient EM signal produced during the timed GW--emitting
phase could be the most efficient way to identify a secure EM
counterpart.

The presence of gas is believed to promote the rapid coalescence of a
SMBH binary (see, e.g., \citealt{escala04,escala05}, but see
\citealt{mm03} and the review by \citealt{gm07} for the possibility of
coalescence without gas, through stellar dynamical processes).  If one
or both SMBHs continue(s) to accrete gas in the GW-emitting stage, for
instance, the coalescing binary may shine as a bright quasar
\citep{koc06,am02} with potential variability on a timescale of $\sim$
hours to days, owing to the inspiraling orbital motion of the binary.
This differs from post-merger counterpart scenarios. For example, the
merger of two nearly equal--mass SMBHs embedded in a thin accretion
disk with a central gap may be expected to produce an X-ray afterglow,
delayed by a disk viscous time relative to the merger, i.e. possibly
$\gsim$ years later~\citep{mp05}. Importantly for prompt counterpart
scenarios, the power emitted in GWs during the final hours of the
coalescence is very large, exceeding $10^{-4}{\,\rm c}^5/{\rm G} \sim
10^{55}\,{\rm erg~s}^{-1}$.  As a result, any coupling of even a small
fraction of this unprecedentedly large power to the surrounding gas
could produce a luminous counterpart, possibly in the form of a
variable EM source.

Motivated by these considerations, in Paper I we developed a formalism
(which we call the Harmonic Mode Decomposition, or HMD method) to
efficiently compute the time--dependent 3--D localization information
contained in the {\it LISA} data stream. We presented an in-depth
study of the potential for pre-merger localizations over the large
parameter space of potential sources.  In particular, we showed that
using the GW inspiral signal accumulated up to some look--back time,
$\tf$, preceding the final coalescence, one already has a good partial
knowledge of the sky position and radial distance to the GW source.
Indeed, since the sky position is deduced primarily from the
detector's motion around the Sun, we found that angular positioning
uncertainties are such that often a source can be localized to within
a few square degrees several weeks prior to merger.  These conclusions
have been confirmed in a recent study \citep{lh07} that included the
additional effects of spin precession and parameter
cross--correlations.  These effects, which are ignored in our work,
impact localization errors only several days or less prior to merger
(resulting in significant improvements at late times).

The main proposal put forward here is to search for an EM counterpart
by monitoring LISA source candidates in the $\sim 2$--$3$ weeks
preceding the merger. We emphasize that in this context, the relevant
criterion is not the comparison between the number density of potential
counterparts and the {\it LISA} error box ($\sim 0.1\deg^2$ in a
typical case), but the much less stringent requirement that the {\it
  LISA} uncertainty be comparable to the field of view (FOV) of a
wide-field survey instrument (say, $\sim 10\deg^2$, such as planned
for the Large Synoptic Survey Telescope, LSST)\footnote{Since the
  redshift of the source can also be inferred from the LISA distance
  measurement, it helps if at least photometric redshifts can be
  measured for all candidates; we will discuss this issue in
  \S~\ref{sec:cutting} below.}. If this requirement is met, it would
be possible to use such an instrument to monitor {\it LISA}
counterpart candidates in real time, during the final stages of the BH
merger.  The goal of the present paper is (i) to discuss,
quantitatively, the feasibility of such a prompt EM counterpart
search, and (ii) to identify the issues that will likely drive the
observational strategies. In particular, we derive source position
uncertainties as a function of time using the results of Paper I, and
investigate survey strategies to best utilize this localization
information.

The remainder of this paper is organized as follows.
In \S~\ref{sec:paper1}, we summarize and expand on the results of Paper I
and present time--dependent {\it LISA} sky position errors.
In \S~\ref{sec:monitoring}, we discuss the possibility of identifying
the EM counterpart of the {\it LISA} source by monitoring the merger
with a wide field astronomical instrument.
In \S~\ref{sec:graviton}, we discuss the additional science that would
be allowed by a real--time EM counterpart identification (as opposed
to identifying a counterpart post--merger).
In \S~\ref{sec:discuss}, we summarize the implications of our findings
and conclude.

\section{Time--Dependent {\it LISA} Sky Position Errors}
\label{sec:paper1}

The first and most fundamental question in searching for prompt EM
counterparts is the accuracy to which the {\it LISA} source can be
localized at various look--back times prior to the coalescence.  In
this section, we present these time--dependent uncertainties.  We use
the Harmonic Mode Decomposition method described in Paper I.  The
reader is referred to that paper for details about the method; here,
for completeness, we summarize the main assumptions before
describing relevant results.

\subsection{Assumptions}
\label{sec:assumptions}

\begin{enumerate}
\item In general, an SMBH inspiral is described by a total of 17
  parameters. These include 2 redshifted mass parameters,\footnote{For
  component masses $m_1$ and $m_2$, the total mass is $M=m_1+m_2$, the
  reduced mass is $\mu=m_1m_2/M$, the symmetric mass ratio is
  $\eta=\mu/M$ and the chirp mass is defined as $\Mchirp=M\eta^{3/5}$
  \citep{Gravitation}, with redshifted masses written as
  $\Mchirp_z=(1+z)\Mchirp$.} $(\Mchirp_z, \eta)$, 6 parameters related
  to the BH spin vectors, ${\bm p}_{\spin}$, the orbital eccentricity,
  $e$, the source luminosity distance, $\dL$, 2 angles locating the
  source on the sky, $(\theta_N, \phi_N)$, 2 angles that describe the
  relative orientation of the binary orbit, $(\theta_{NL},
  \phi_{NL})$, a reference time and a reference phase at merger,
  i.e. at the innermost stable circular orbit (ISCO), $t_{\rm ISCO}$,
  $\phi_{\rm ISCO}$, and, finally, the instantaneous orbital phase,
  $\phi_{\rm orb}$ (which specifies the time-evolution of the binary).
  We denote the lookback time before merger by $\tf$. Throughout the
  paper, all time parameters will be quoted in redshifted units
  (i.e. as measured by an observer on Earth).
\item The signal for a GW inspiral is determined by the above set of
  parameters and two additional angles describing the orientation of
  the {\it LISA} constellation, $(\Xi,\Phi)$.  Note that these angles
  precess with time; indeed, this time--dependence carries all of the
  sky localization information.  Since the period is precisely equal
  to {\it LISA}'s orbital period of one year, we need to specify only
  some reference values (e.g. chosen at the time of the merger,
  $\Xi_{\rm ISCO}$ and $\Phi_{\rm ISCO}$).
\item We use the circular, restricted post-Newtonian (PN)
  approximation for the GW waveform \citep{cf94} keeping only the
  quadrupole GW harmonic for the GW amplitude. Higher order GW
  harmonics are negligible if the quadrupole harmonic is within the
  LISA-frequency band for at least a few months, but can improve
  uncertainties for large mass binaries with $M_z (\eta/0.25)^{3/5}
  \gsim 2\times 10^7\Msun$ and much shorter observation times
  \citep{hm03,Arun07a,ts07}.
\item We neglect the effects of Doppler phase modulation due to {\it
LISA}'s orbital motion which is an excellent approximation for SMBH
inspirals (S. A. Hughes, private communication, 2006;
\citealt{Arun07b}).
\item We neglect SMBH spins and, in particular, neglect the
  spin--orbit precession for determinations of the angular
  parameters. This assumption is useful in simplifying our equations
  and in focusing on the behavior of pure angular
  modulation. Including the effect of spin helps reduce uncertainties
  in the final days and hours of the coalescence \citep{lh07}.

\item We assume circular orbits ($e=0$). Although eccentricity is
  efficiently damped by gravitational radiation reaction
  \citep{Peters64}, the presence of gaseous circumbinary disks could
  lead to non-zero eccentricities for at least some {\it LISA}
  inspiral events \citep{ArmNat05,PapNelMas01,macfadyen07}.  For
  highly eccentric orbits, higher order harmonics appear in the GW
  phase, which could again lead to improved errors.
\item We fix the start of {\it LISA} observations at a look--back time
$\ti\equiv \min\{t(f_{\min}),1\yr\}$ prior to merger.\footnote{The
look--back time throughout this paper is defined to run ``backwards''
from ISCO, i.e. we set $t_{\rm merger}=0$ (see Paper I).}  This
corresponds to the time when the GW inspiral frequency $f$ crosses the
low frequency noise wall of the detector at $f_{\min}=0.03\mHz$, but
we limit the initial look--back time to a maximum of $1\yr$ before
merger.
\item We assume that the slow annual amplitude modulation due to {\it
    LISA}'s orbital motion can be used to determine the luminosity
    distance and angular parameters, ${\bm
    p}_{\slow}=\{\dL,\theta_N,\phi_N,\theta_{NL},\phi_{NL}\}$, while
    the other remaining parameters, ${\bm
    p}_{\fast}=\{\Mchirp_z,\mu_z,t_{\rm ISCO},\phi_{\rm ISCO}\}$, are
    determined independently from the high frequency GW phase, with
    negligible cross-correlation between the two sets of parameters.
    This has been verified to be a good assumption up to a few days
    prior to ISCO, with cross--correlations increasing the
    localization errors only by a few tens of $\%$. Closer to merger,
    however, cross-correlations can increase final errors at ISCO by a
    factor of 2-3 \citep{lh07}.
  \item We follow \citet{bc04} in calculating the spectral noise
    density, $S_{n}(f)$, which includes the instrumental noise as well
    as galactic/extra-galactic backgrounds. For the instrumental
    noise, we use the effective non-angular averaged online {\it
      LISA} Sensitivity Curve Generator\footnote{{\tt
        www.srl.caltech.edu/$\sim$shane/sensitivity/}} \citep{bbw05},
    while we use the isotropic formulae for the galactic and
    extra-galactic background \citep{bc04}. We do not consider errors
    from possible theoretical uncertainties \citep{vc07}, which are
    negligible at look-back times $\tf$ well before $t_{\rm ISCO}$,
    when the binary separations $r\gg r_{\rm ISCO}$.
\item We estimate the expected parameter uncertainties using the
  Fisher matrix approach, using the Harmonic Mode Decomposition (HMD)
  method developed in Paper I (essentially a discrete Fourier
  transform of the GW waveform with fundamental frequency yr$^{-1}$
  corresponding to the orbital period of the detector).  The main
  advantage of the HMD method is that the time--dependence of errors
  can be extracted independently of the specific SMBH binary orbital
  elements, reducing the computation time by many orders of
  magnitude. This allows us to survey simultaneously the dependencies
  on source sky position, SMBH masses and redshift.  We carry out
  Monte--Carlo (MC) calculations with $2\times 10^4$ random samples
  for the angles $\cos \theta_{N},\cos\theta_{NL},\phi_{NL}$, and
  $\alpha,\Phi_{\rm ISCO}$.  Here $\alpha$ is a time-independent
  combination of $\Phi(t)$ and $\Xi(t)$, defined as $\alpha \equiv
  \Xi-\Phi +\phi_{N} - \frac{3\pi}{4}$.  The angles are defined
  relative to LISA's orbital plane (i.e., relative to the ecliptic),
  with $\theta_N$ and $\theta_{NL}$ referring to polar and $\phi_N$
  and $\phi_{NL}$ referring to azimuthal coordinates.  Several
  thousand values of $M$ and $z$ are considered, in the range
  $10^5<M/\Msun<10^8$ and $0.1\leq z\leq7$.  In addition, we ran
  specific MC calculations to study possible systematic effects with
  respect to the source sky position (on a grid of several hundred
  values).

\end{enumerate}

\subsection{Evolution of Error Ellipse}
\label{sec:errors}

\subsubsection{Size of the Error Ellipse}

\begin{figure*}[tb]
\centering
\mbox{\includegraphics[height=5.0cm]{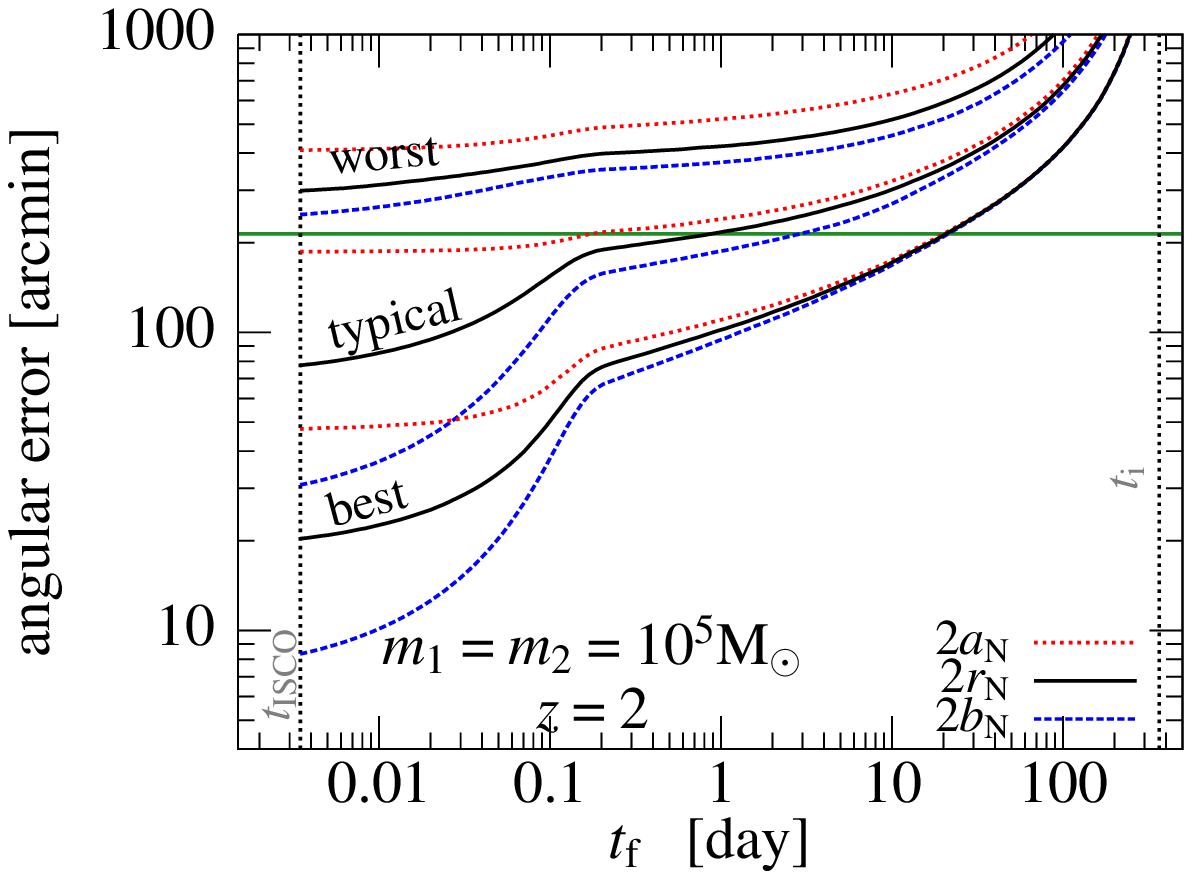}}
\mbox{\includegraphics[height=4.85cm]{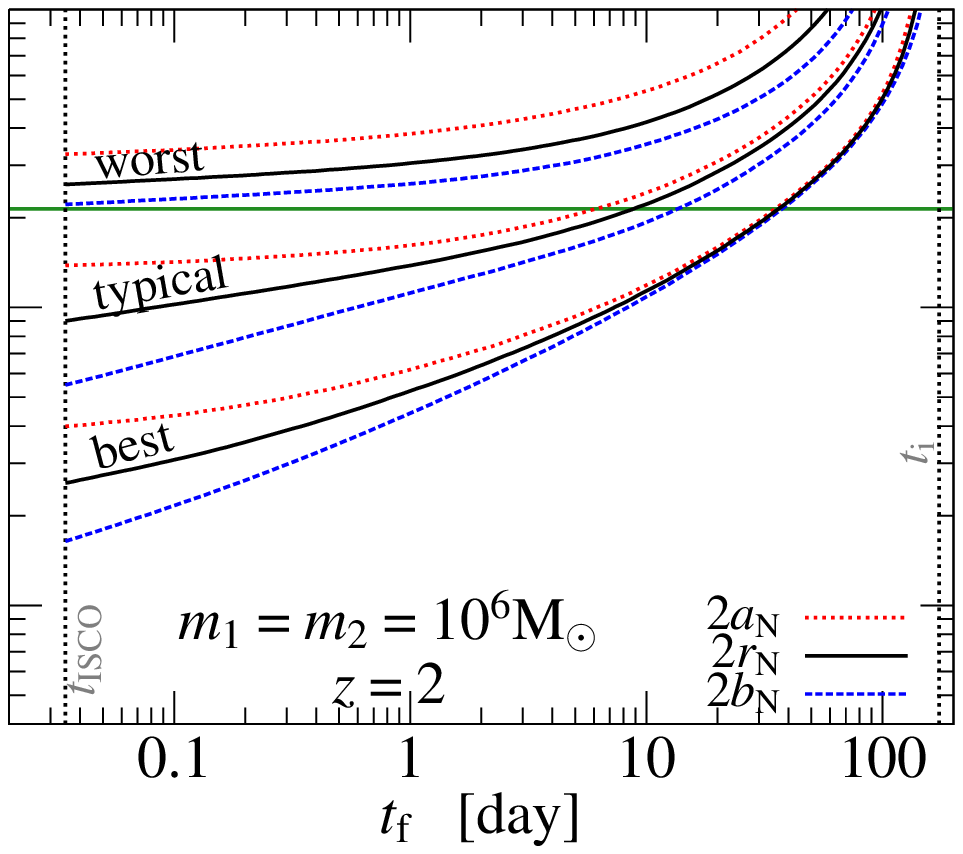}}
\mbox{\includegraphics[height=4.85cm]{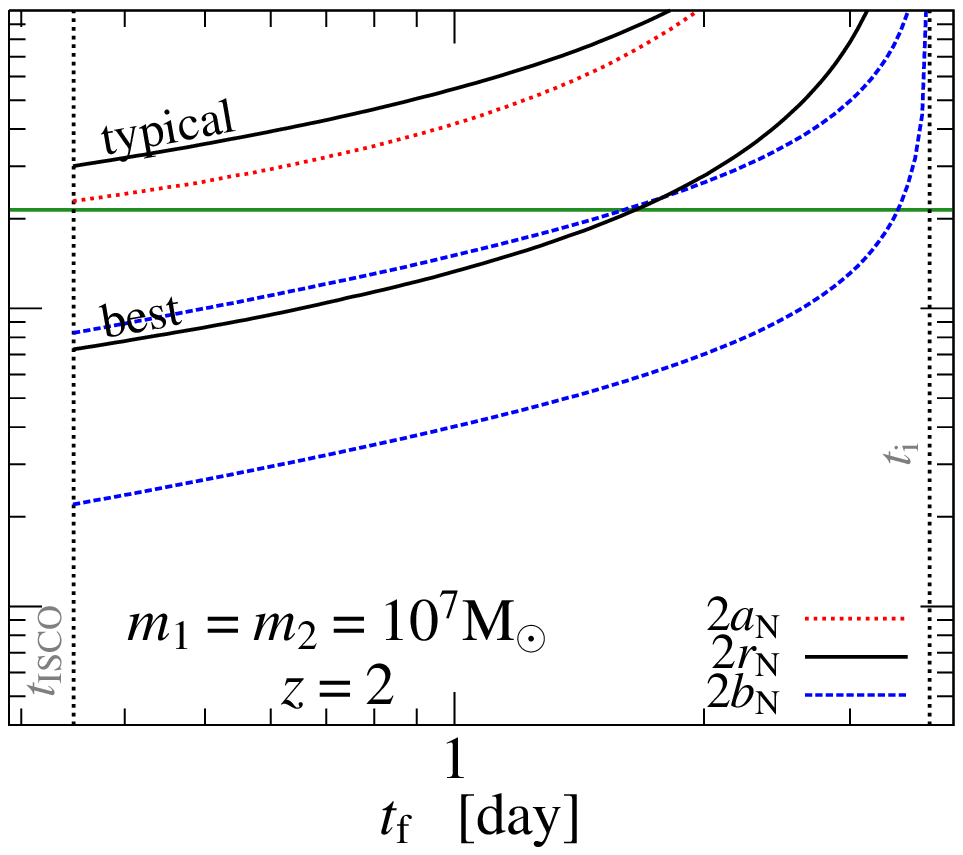}}\\
\mbox{\includegraphics[height=5.0cm]{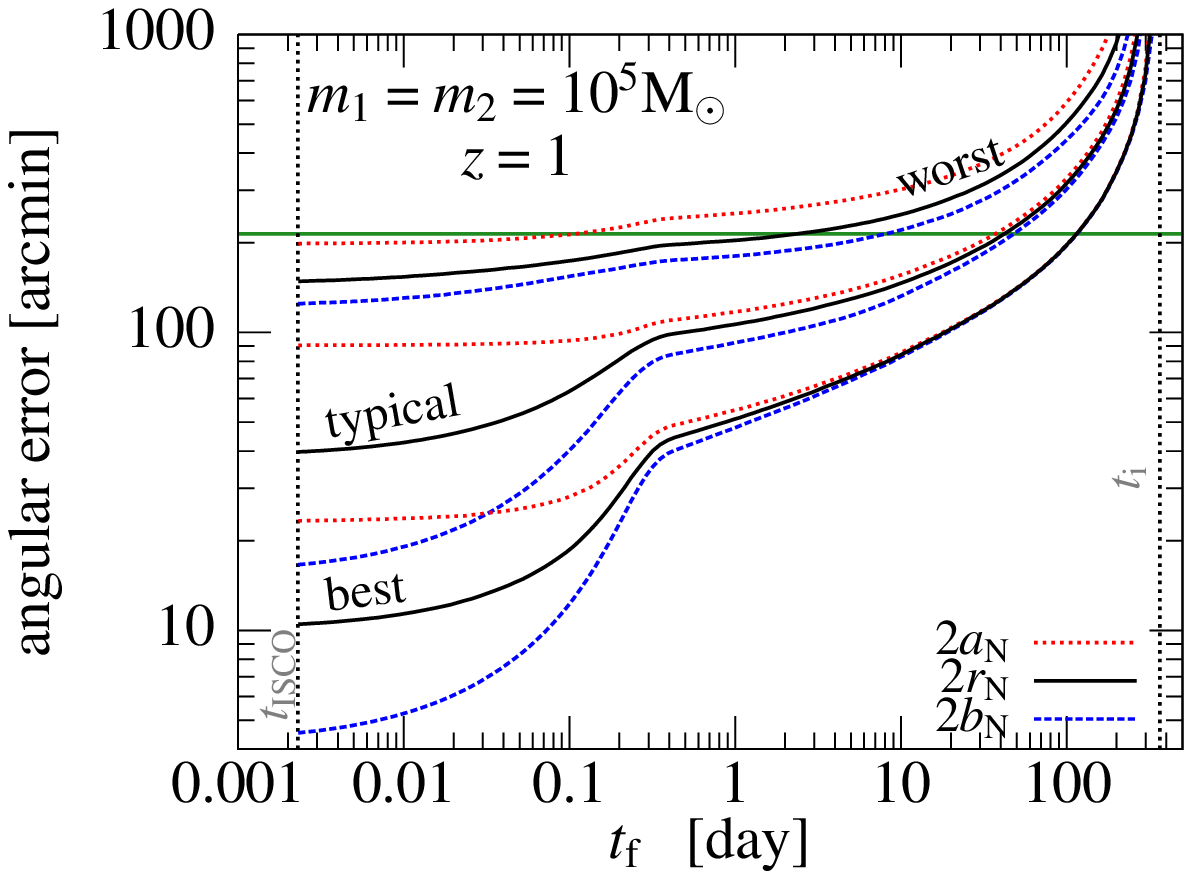}}
\mbox{\includegraphics[height=4.85cm]{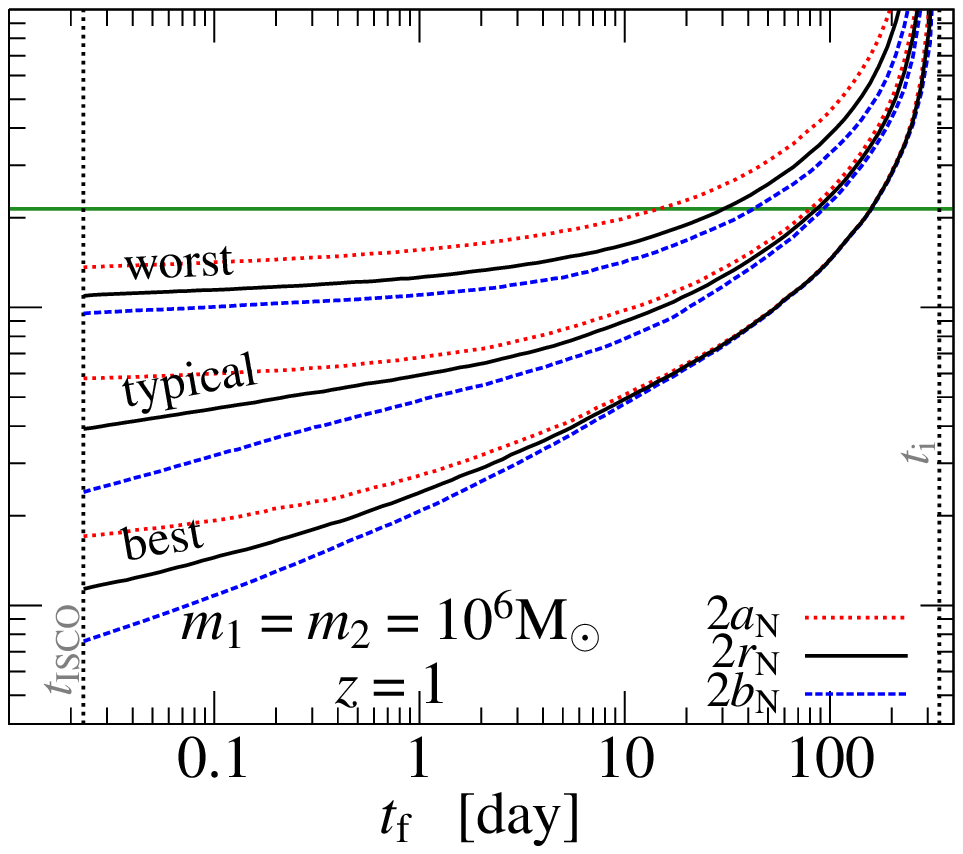}}
\mbox{\includegraphics[height=4.85cm]{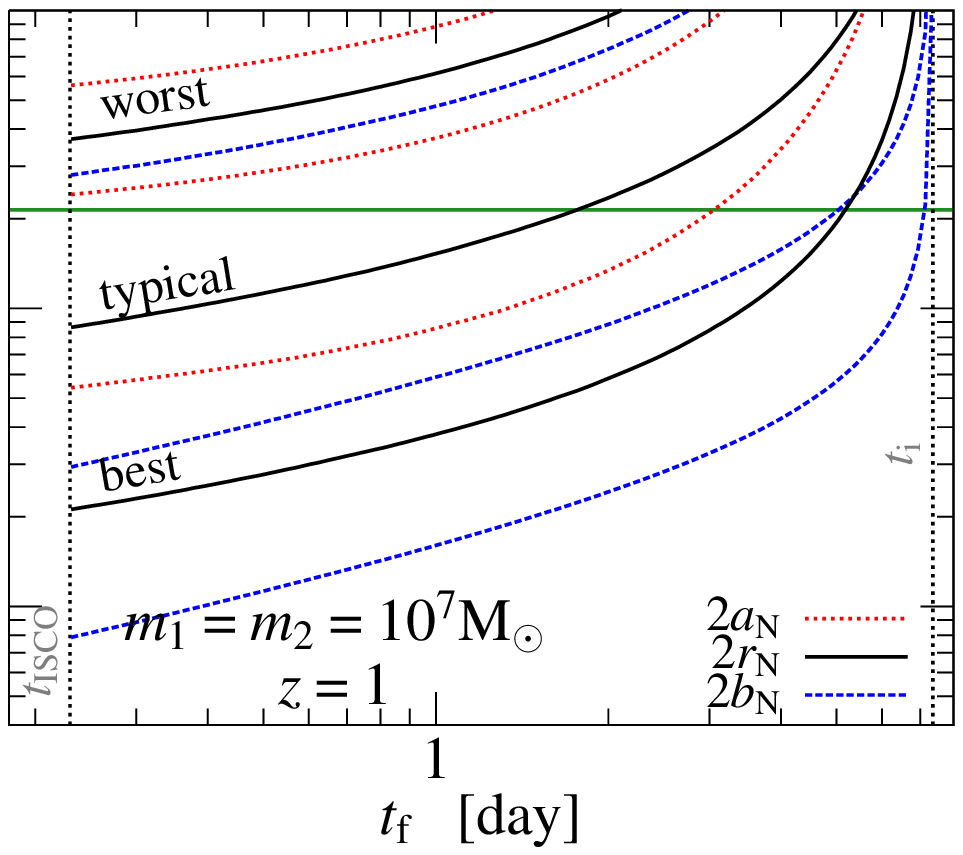}}
\caption{\label{f:errortau} Evolution with pre-ISCO look--back time,
  $\tf$, of LISA source localization errors for the sky position
  (major axis $2a$, minor axis $2b$, and equivalent diameter,
  $2r\equiv 2\sqrt{ab}$, of the error ellipsoid). Six cases with
  $M=2\times 10^{5,6,7}\Msun$ and $z=1,2$ are shown as labeled.  Best,
  typical, and worst cases for random orientation events represent the
  $10\%$, $50\%$, and $90\%$ levels of cumulative error distributions,
  respectively.  The left panels show that the eccentricity of the
  angular error ellipsoid can change significantly during the last
  day.  The absolute errors are typically small enough to fit in the
  FOV of a wide--field instrument in the last 1-2 weeks.  The
  horizontal lines show a diameter of $3.57^{\circ}$, which
  corresponds to localizing the source to within 10 deg$^2$.
}
\end{figure*}

With the above assumptions, we obtain the two--dimensional sky
position error ellipses.  The time--evolving $1\sigma$ errors on the
size of the ellipse are shown for a few examples in
Figure~\ref{f:errortau}.\footnote{Note that throughout this paper, we
  show $1\sigma$ errors. Higher confidence levels will obviously
  require covering proportionately larger areas.}  We also note that
the Fisher matrix approach, by construction, approximates the sky
position errors (and all other parameter uncertainties) by
ellipses. This approximation has been shown for the $1\sigma$
uncertainties to be in excellent agreement with the posterior
probability distributions for the slow parameters determined from
Monte Carlo Markov Chain and Bayesian methods \citep{cp06,val07}.
Figure~\ref{f:errortau} shows the gradual improvement in localization,
for SMBH coalescences with six different fixed mass and redshift
combinations, in the form of the major axis ($2a$), minor axis ($2b$)
and equivalent diameter ($2r=\sqrt{4ab}$).

Figure~\ref{f:errortau} displays results for three separate cumulative
probability distribution levels, $90\%,50\%,10\%$, so that $10\%$
refers to the best $10\%$ of all events, as sampled by the random
distribution of the five angular parameters.  Note that the levels
refer to the parameter in question (i.e. the 10\% of all events for
which $a$ is best--measured, in general, is different from the 10\% of
the events for which $b$ is best measured), so eccentricities can not
be directly read off from this figure.
The evolution of errors scales steeply with look--back time for
$\tf\gsim 40\dy$. For smaller look--back times, errors essentially
stop improving in the ``worst'' ($90\%$ level) case, improve with a
relatively shallow slope for the ``typical'' ($50\%$ level) case, and
improve more steeply in the ``best'' case ($10\%$ level among the
realizations of fiducial angular parameters).  These behaviors are
explained in Paper I; the minor-axis best case approaches the simple
$(S/N)^{-1}$ scaling expected in the absence of parameter
degeneracies.

\begin{figure*}[tbh]
\centering
\mbox{\includegraphics[width=6.4cm]{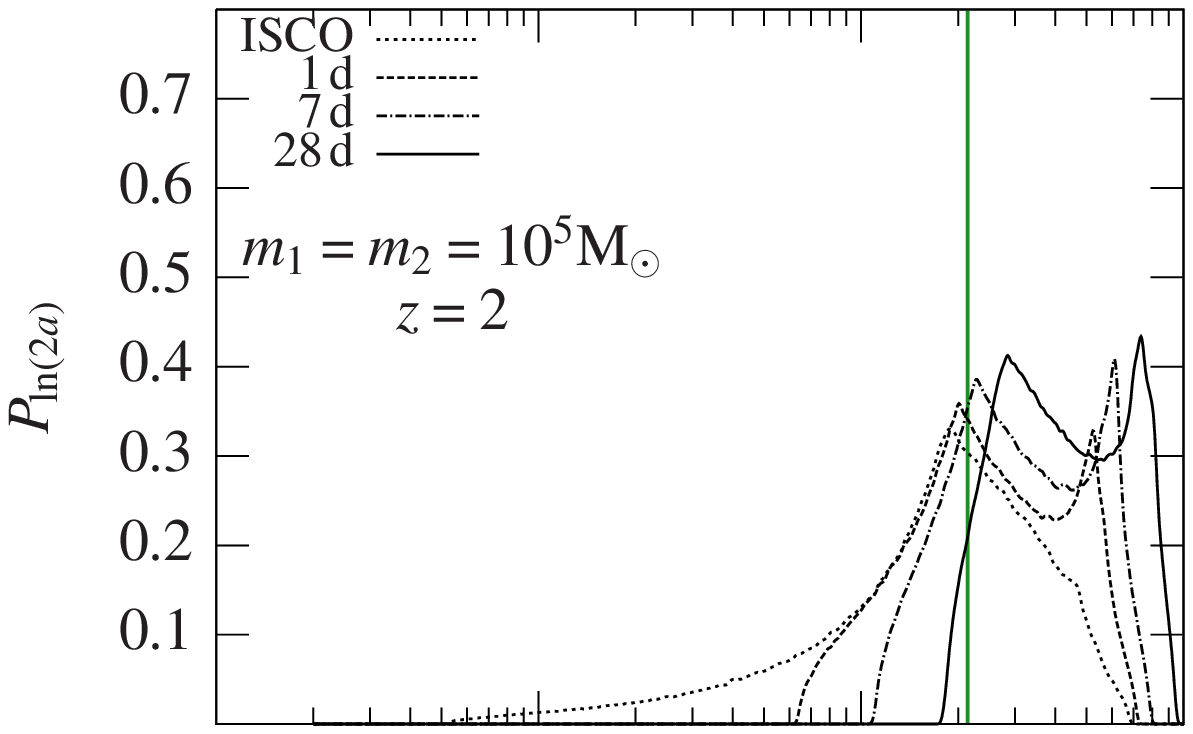}}
\mbox{\includegraphics[width=5.3cm]{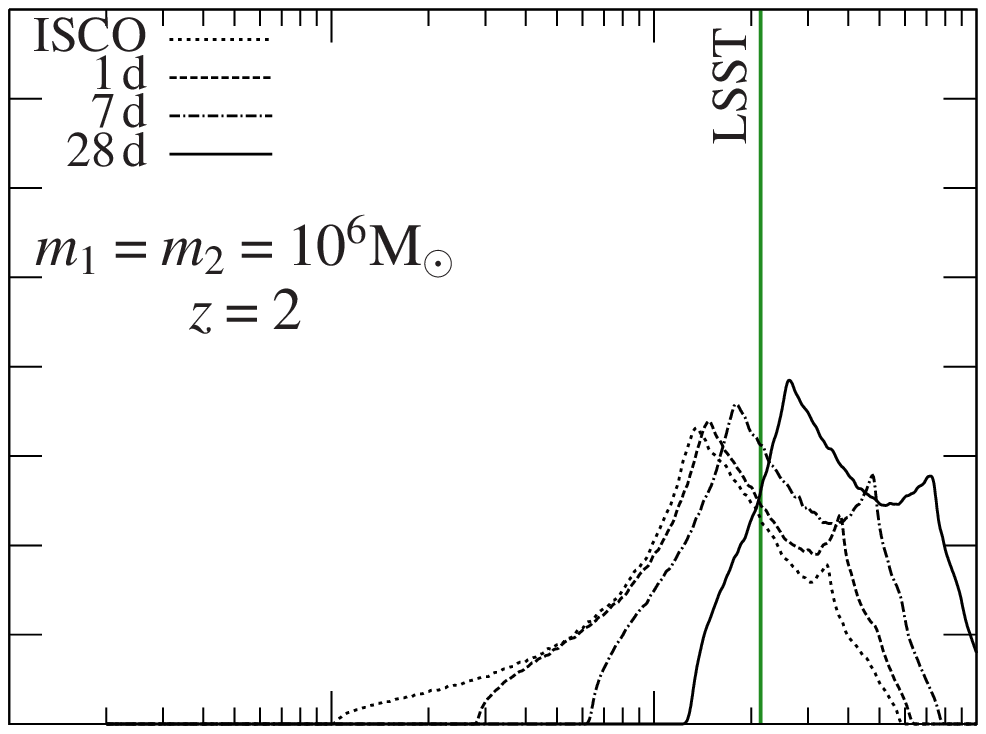}}
\mbox{\includegraphics[width=5.3cm]{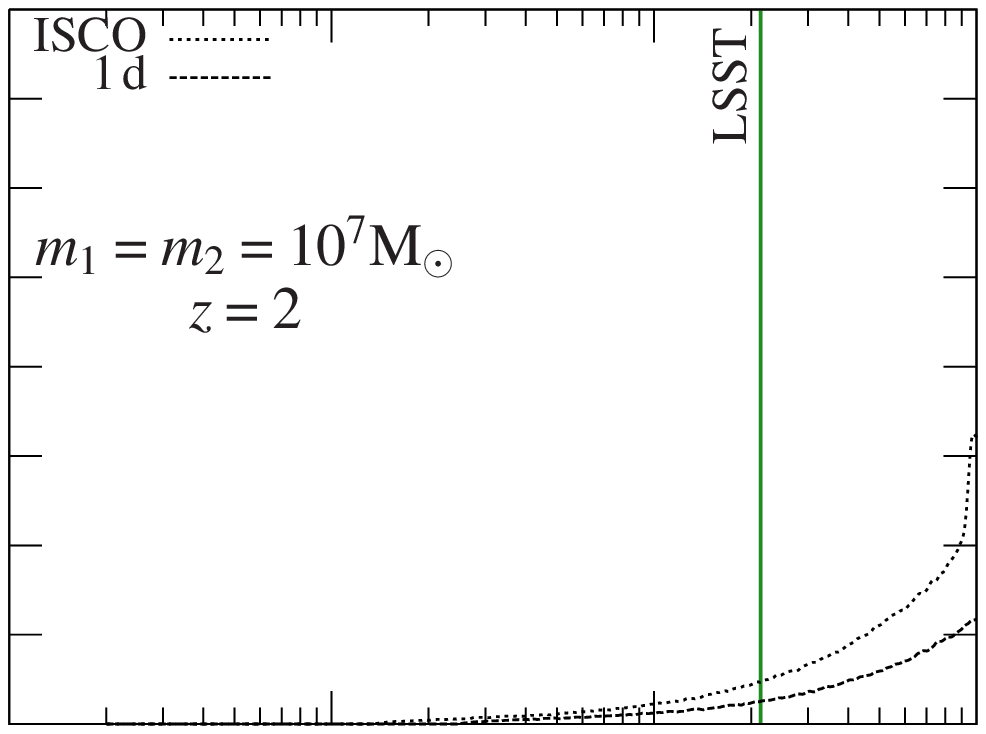}}\\
\mbox{\includegraphics[width=6.4cm]{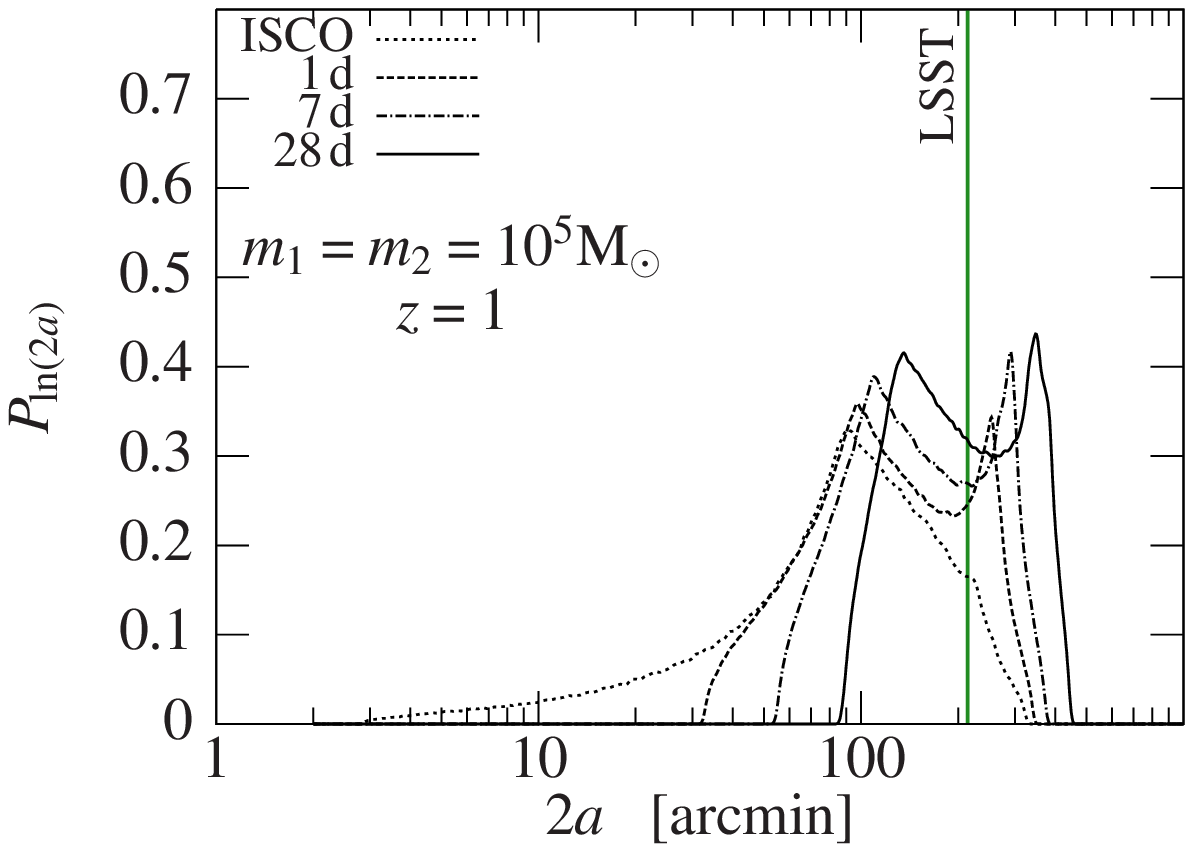}}
\mbox{\includegraphics[width=5.3cm]{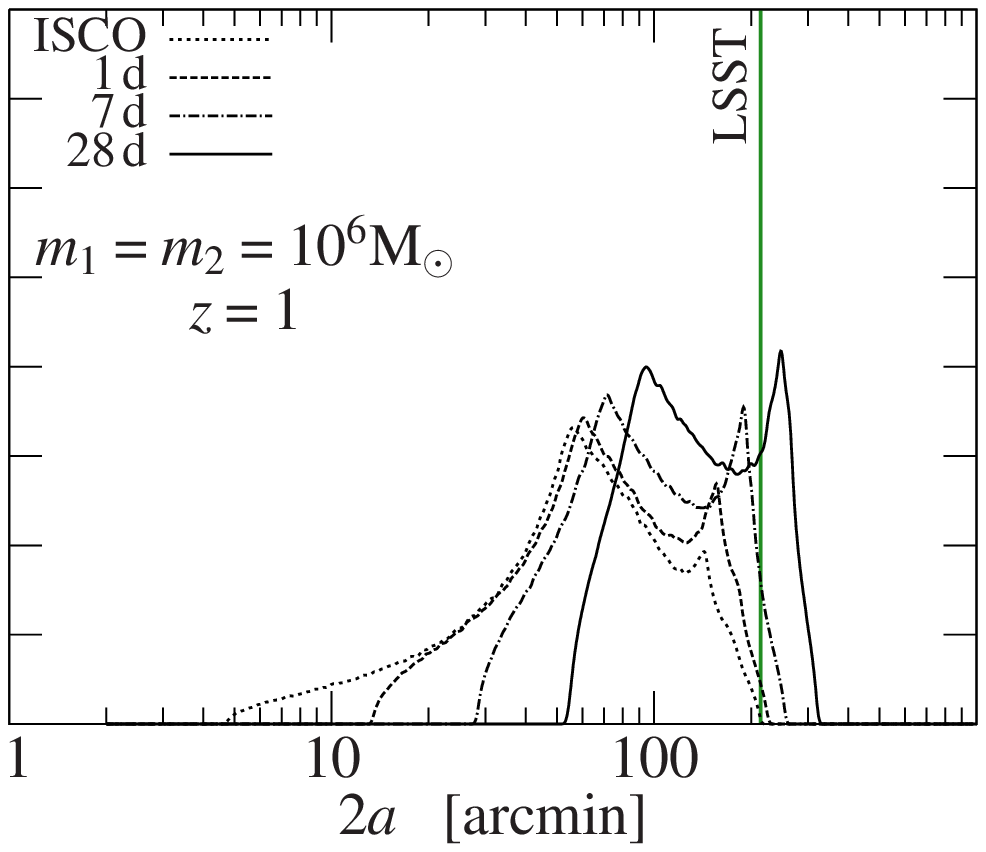}}
\mbox{\includegraphics[width=5.3cm]{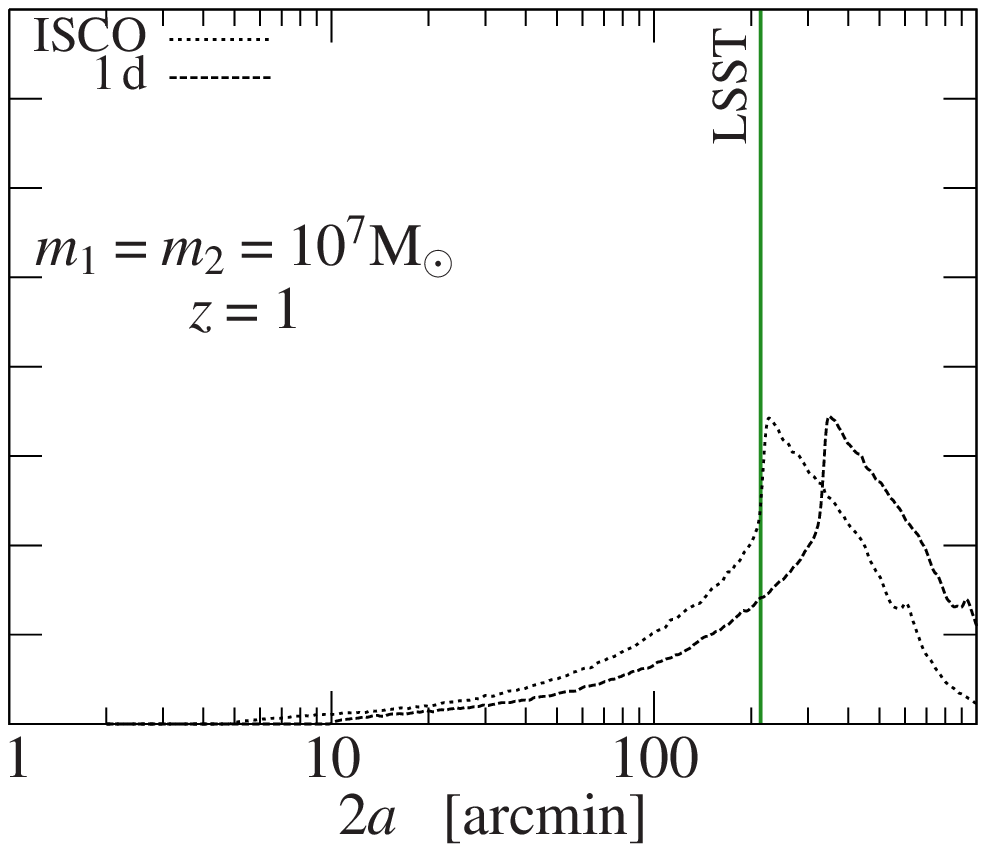}}
 \caption{\label{f:snapshot_a}
Probability density of the $1\sigma$ LISA source localization ellipsoid major axis at 4 snapshots: ISCO, and 1, 7, and 28 days before merger. Each curve shows the distribution in logarithmic bins, $\ln(2a)$, calculated using $10^6$ random binaries. The vertical line shows the FOV of LSST for comparison.
}
\end{figure*}
\begin{figure*}[tbh]
\centering
\mbox{\includegraphics[width=6.4cm]{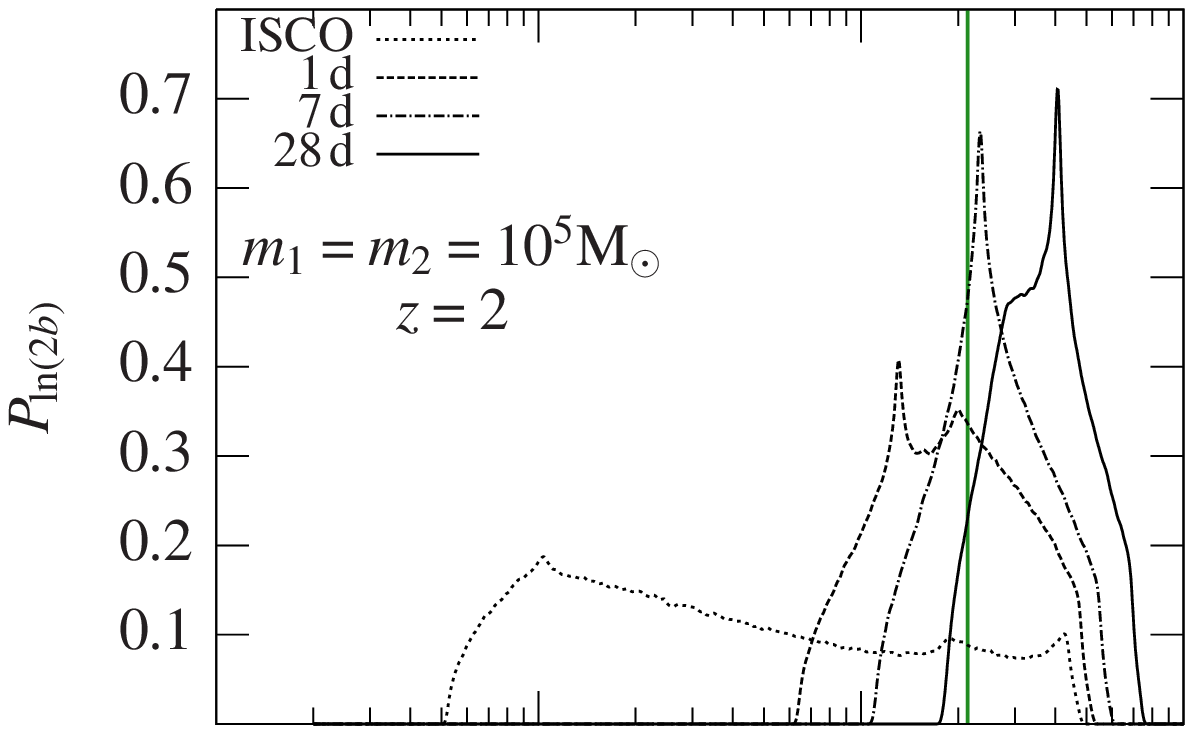}}
\mbox{\includegraphics[width=5.3cm]{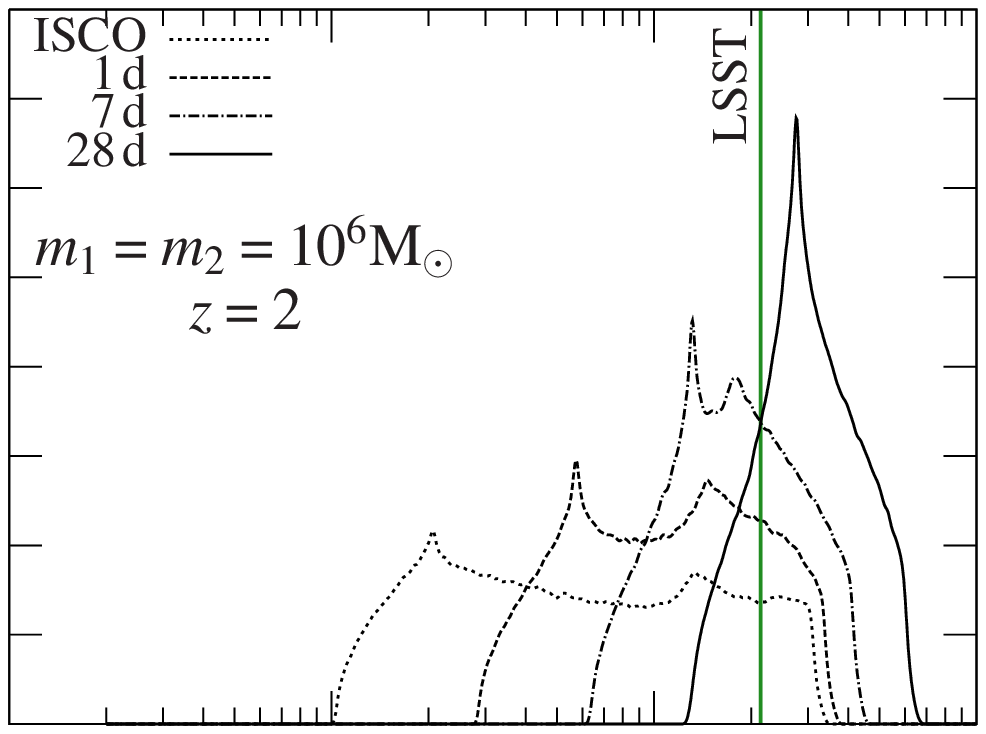}}
\mbox{\includegraphics[width=5.3cm]{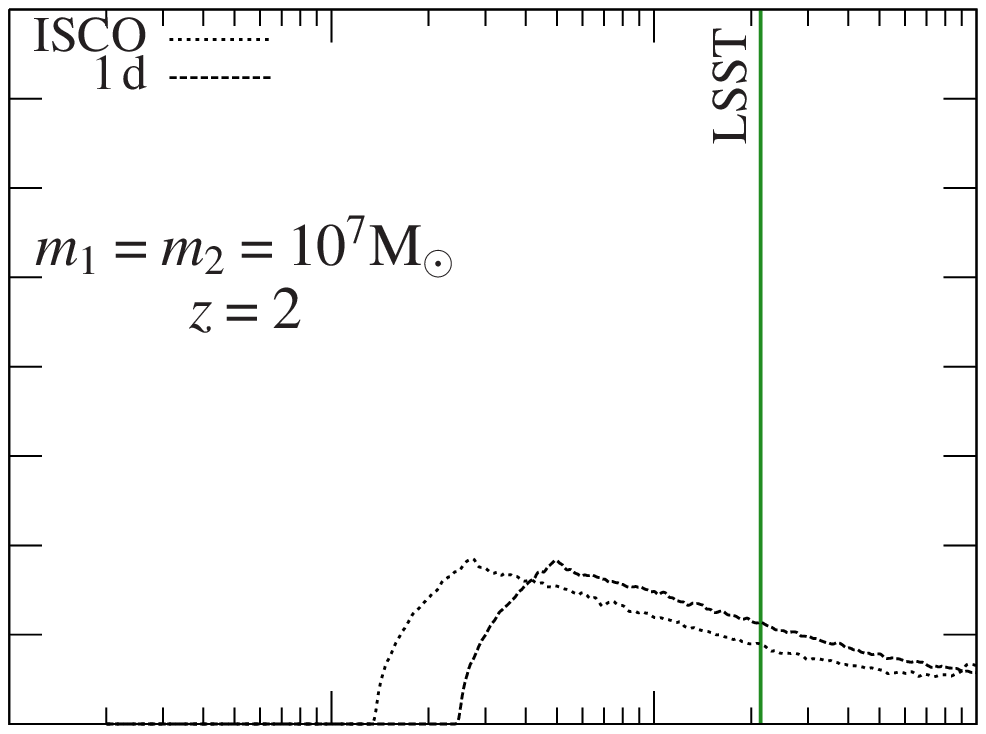}}\\
\mbox{\includegraphics[width=6.4cm]{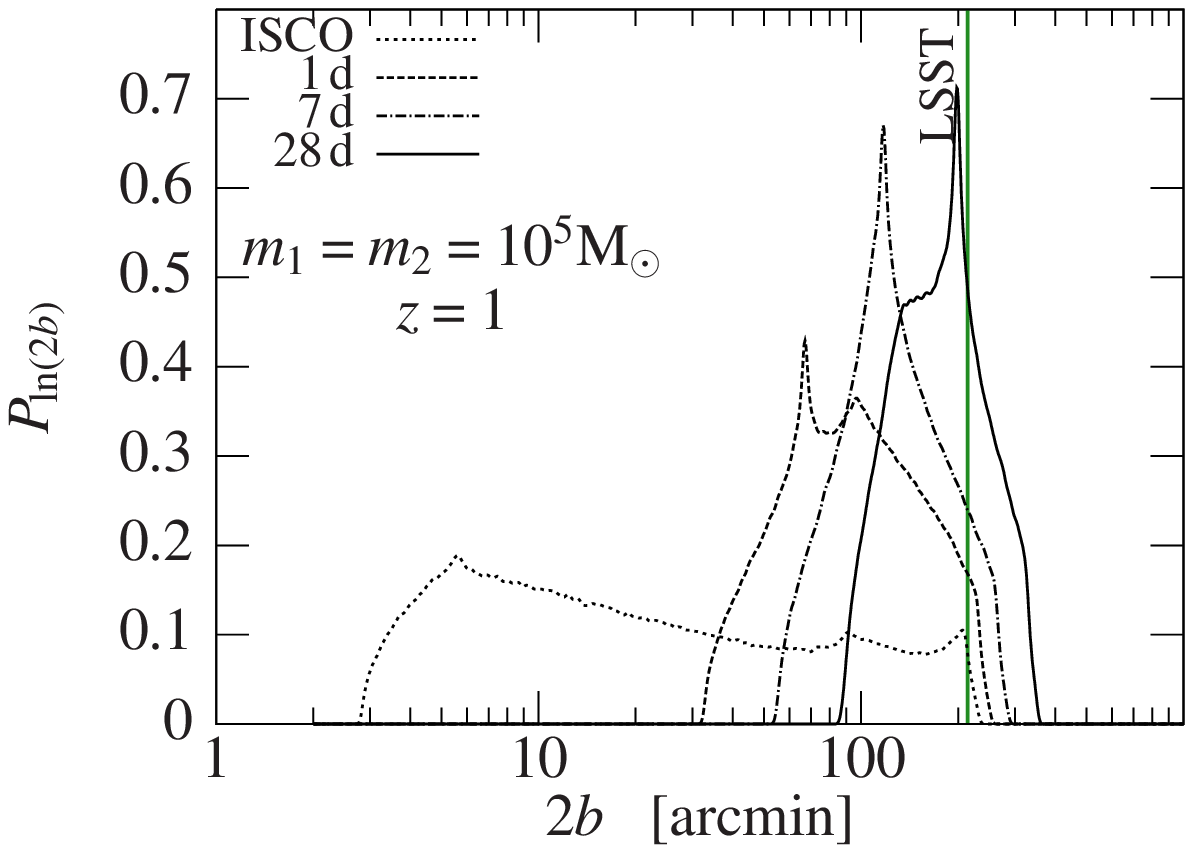}}
\mbox{\includegraphics[width=5.3cm]{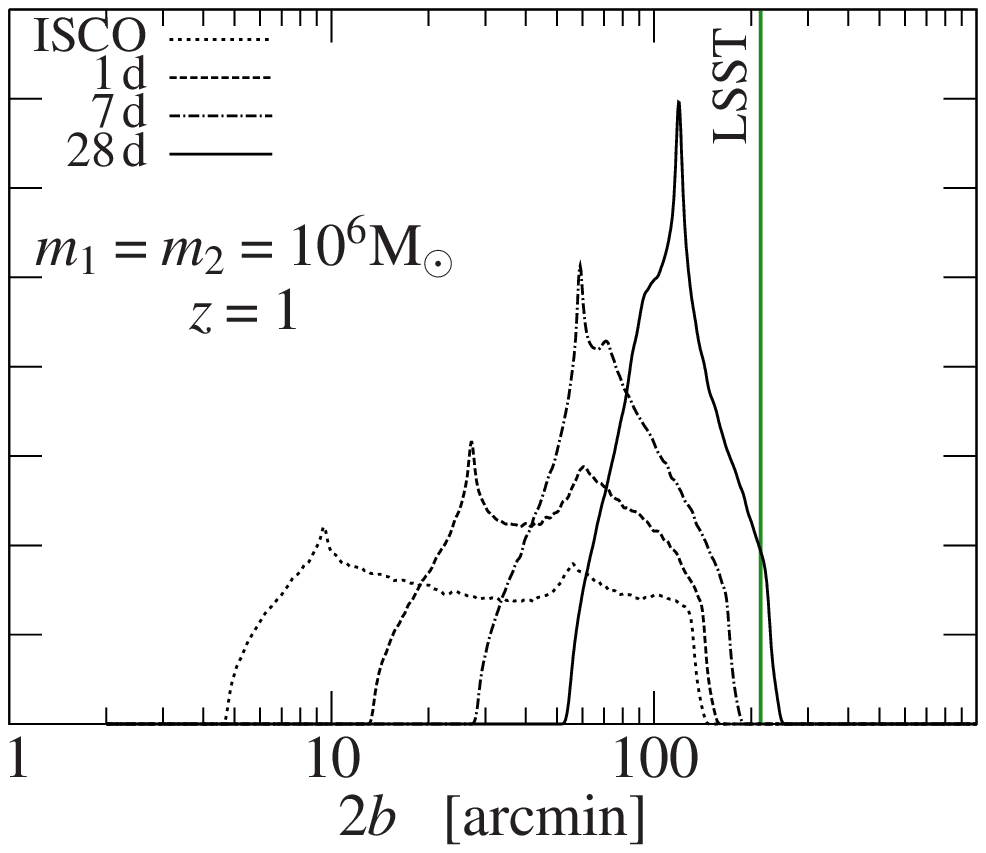}}
\mbox{\includegraphics[width=5.3cm]{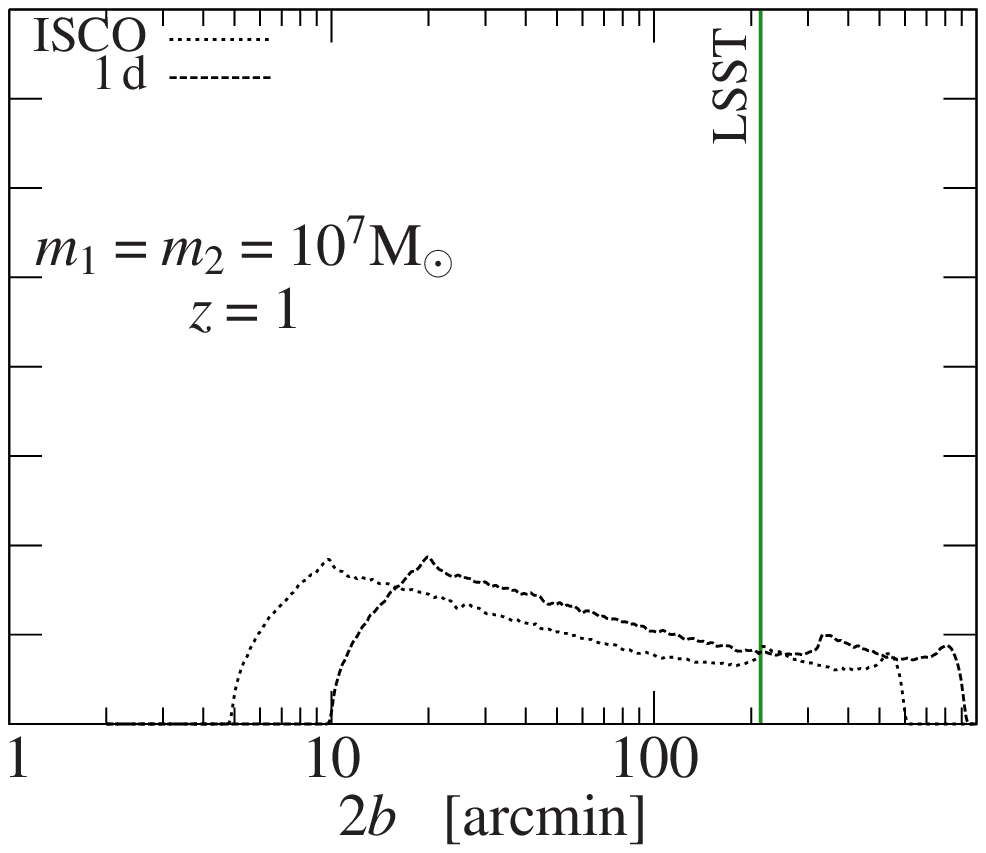}}
 \caption{\label{f:snapshot_b}
Same as Figure~\ref{f:snapshot_a} but for the sky localization ellipsoid minor axis.
}
\end{figure*}

In Figures~\ref{f:snapshot_a} and \ref{f:snapshot_b}, we show the
probability distribution for the errors $2a$ and $2b$ in logarithmic
bins at 4 different snapshots: at ISCO, and at 1, 7, and 28 days
before merger, using $10^6$ random binaries.  These distributions are
broad, spanning 1--2 orders of magnitude. They are generally
asymmetric, skewed right or left for the major and minor axis,
respectively.  The distribution evolves with look--back time: it is
initially peaked, gets flatter and broader approaching merger, and
develops a bimodal structure with two peaks. This bimodal nature of
the distribution at ISCO is also visible in the histograms of
\citet{lh07}, which our Figures~\ref{f:snapshot_a} and
\ref{f:snapshot_b} confirm at high resolution.  The wide distributions
suggest a non--negligible chance of being lucky/unlucky and finding a
very well/poorly localized source.

\subsubsection{Eccentricity of the Error Ellipse}

Figures~\ref{f:errortau}, \ref{f:snapshot_a} and \ref{f:snapshot_b}
suggest that the 2D sky localization error ellipsoid is initially
circular ($a\approx b$), but the geometry changes significantly during
the last stages of the merger. For example, in the typical case (for
$z=1$ and $m_1=m_2=10^6~{\rm M_\odot}$), in the last two weeks, the
major axis improves much more slowly than the minor axis.  The effect
is much more pronounced at lower masses. As the left panels show, the
major axis essentially stops improving within the last $\sim$day,
while the minor axis maintains a steep evolution, so that the
eccentricity of the 2D angular error ellipsoid changes rapidly (in
fact this behavior is even more pronounced if spin precession effects
are included, \citealt{lh07}).  The fact that the error ellipse, in
general, changes its eccentricity and its orientation may need to be
taken into account in observational strategies (although, of course,
as time evolves, subsequent error ellipsoids always lie entirely
within the earlier error ellipsoids).

\begin{figure*}[tbh]
\centering
\mbox{\includegraphics[height=4.85cm]{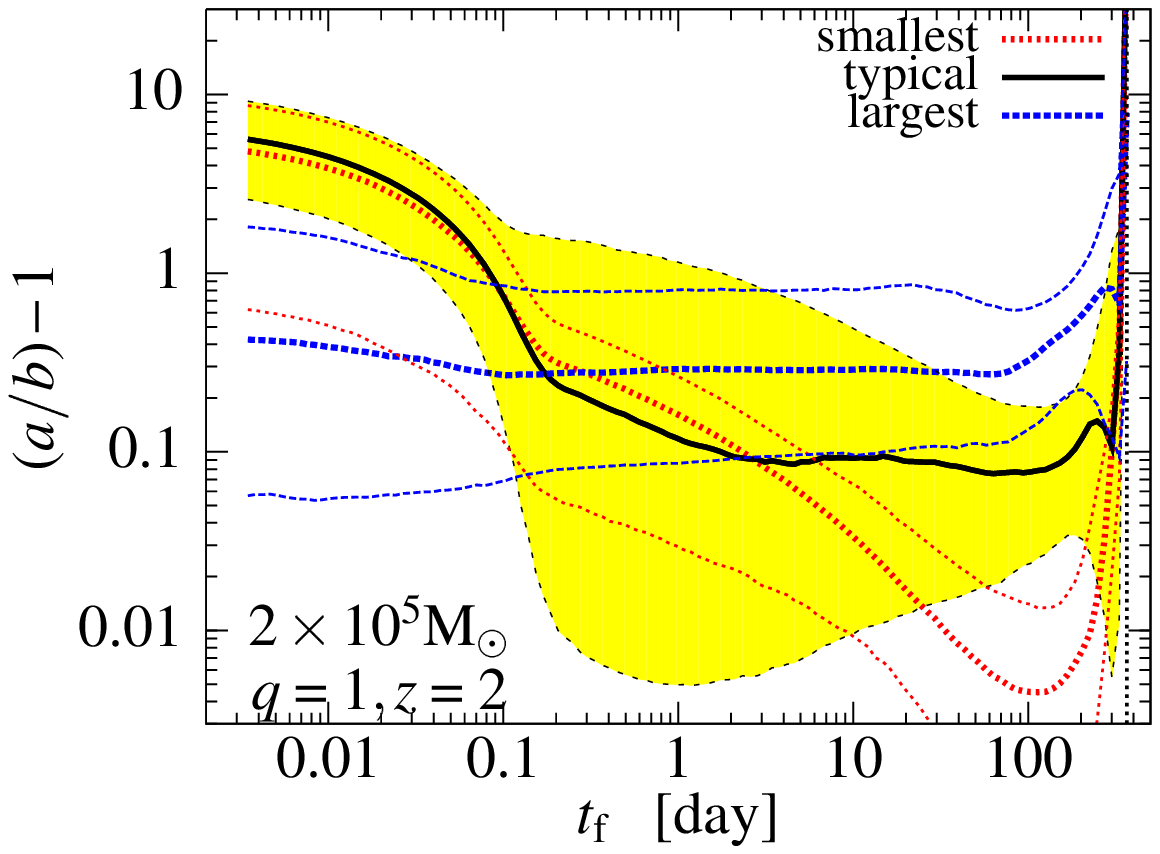}}
\mbox{\includegraphics[height=4.85cm]{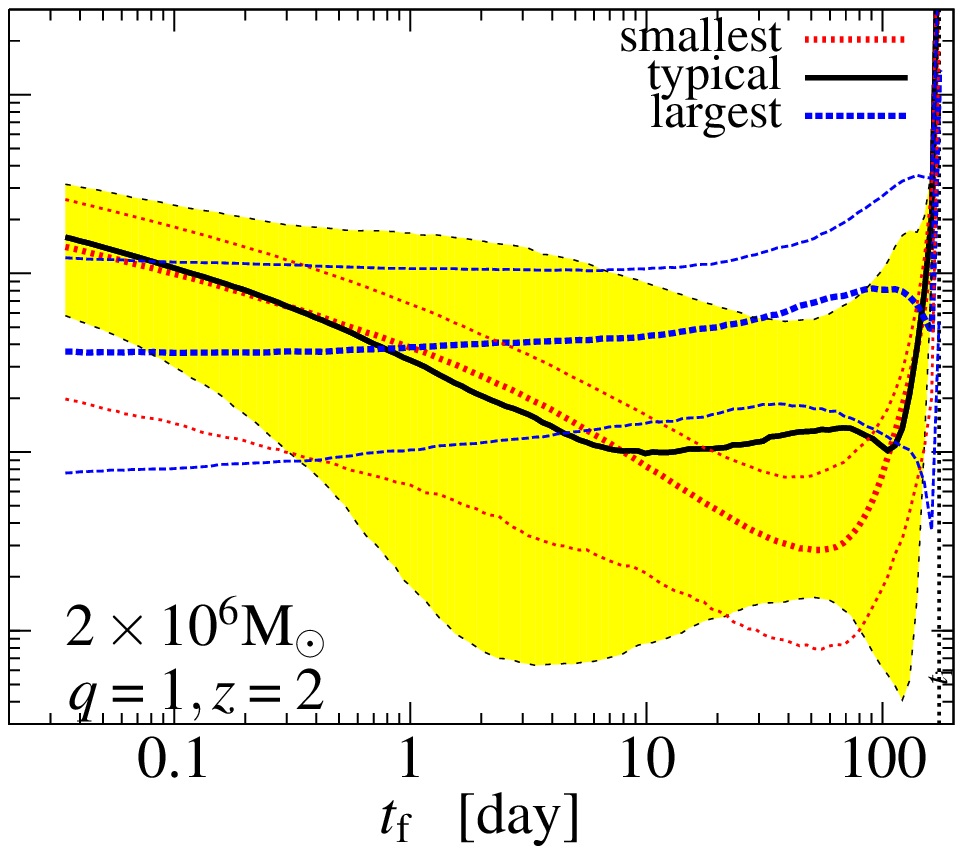}}
\mbox{\includegraphics[height=4.85cm]{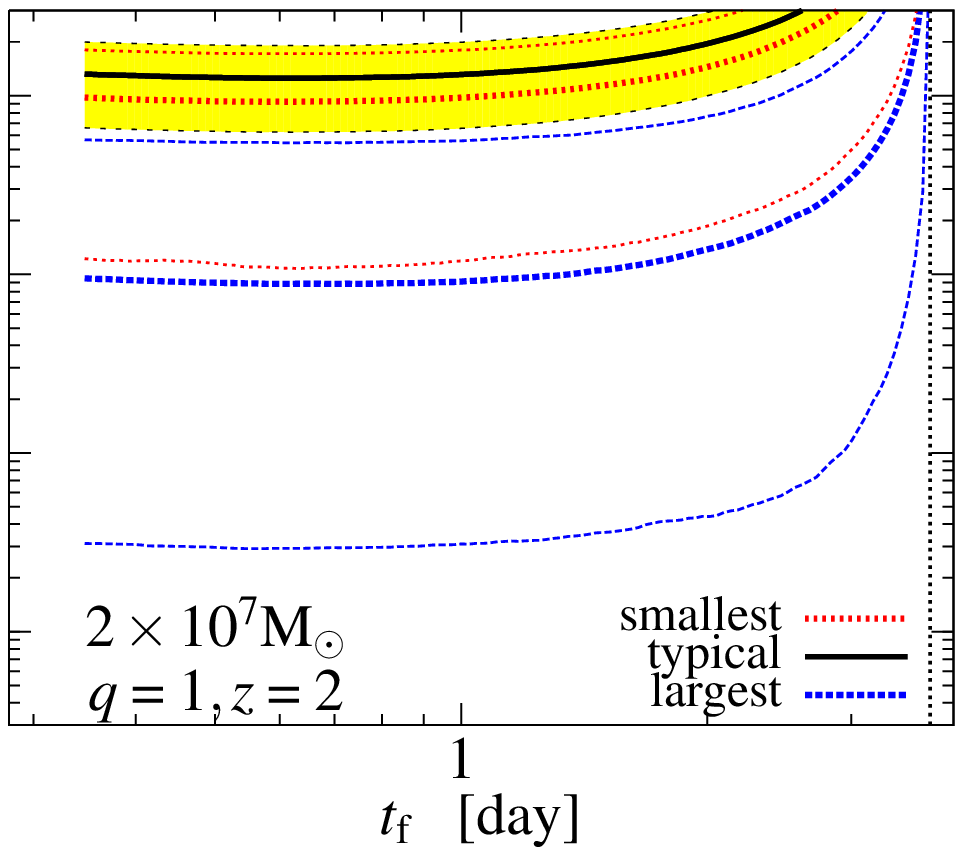}}\\
\mbox{\includegraphics[height=4.85cm]{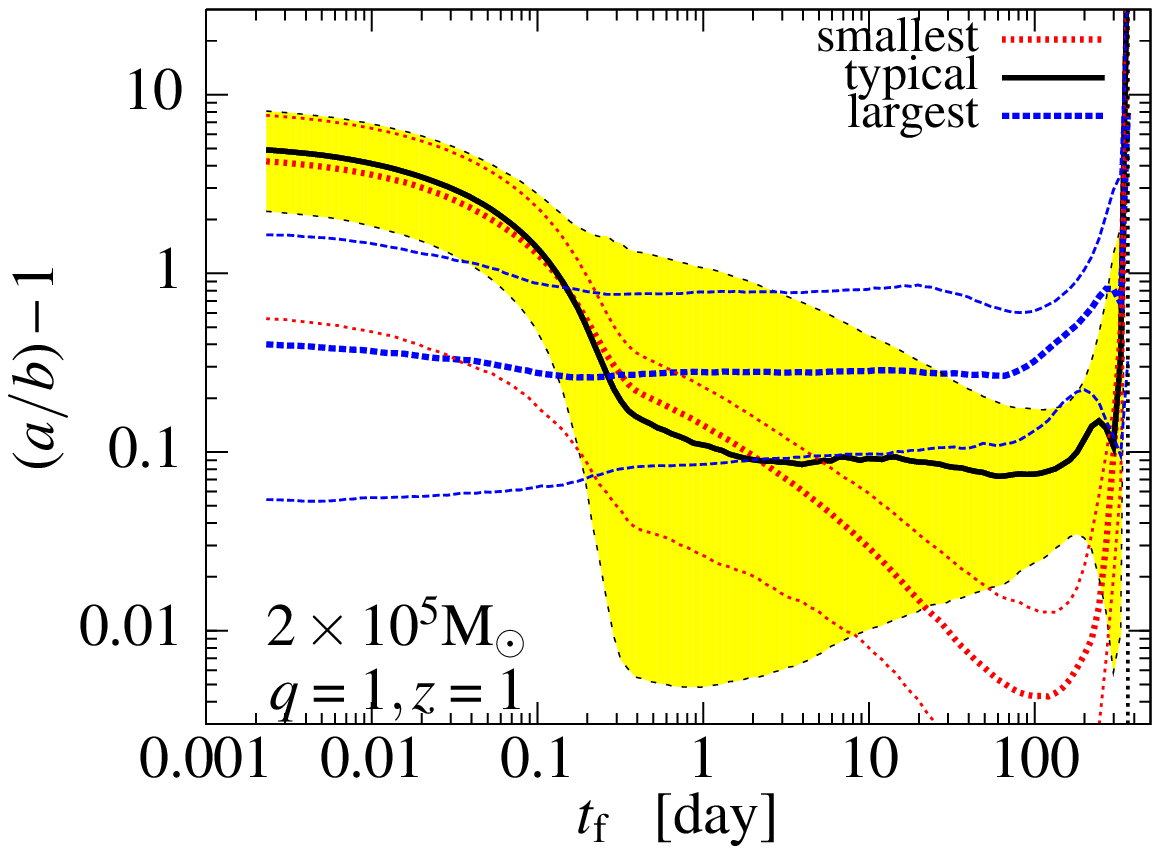}}
\mbox{\includegraphics[height=4.85cm]{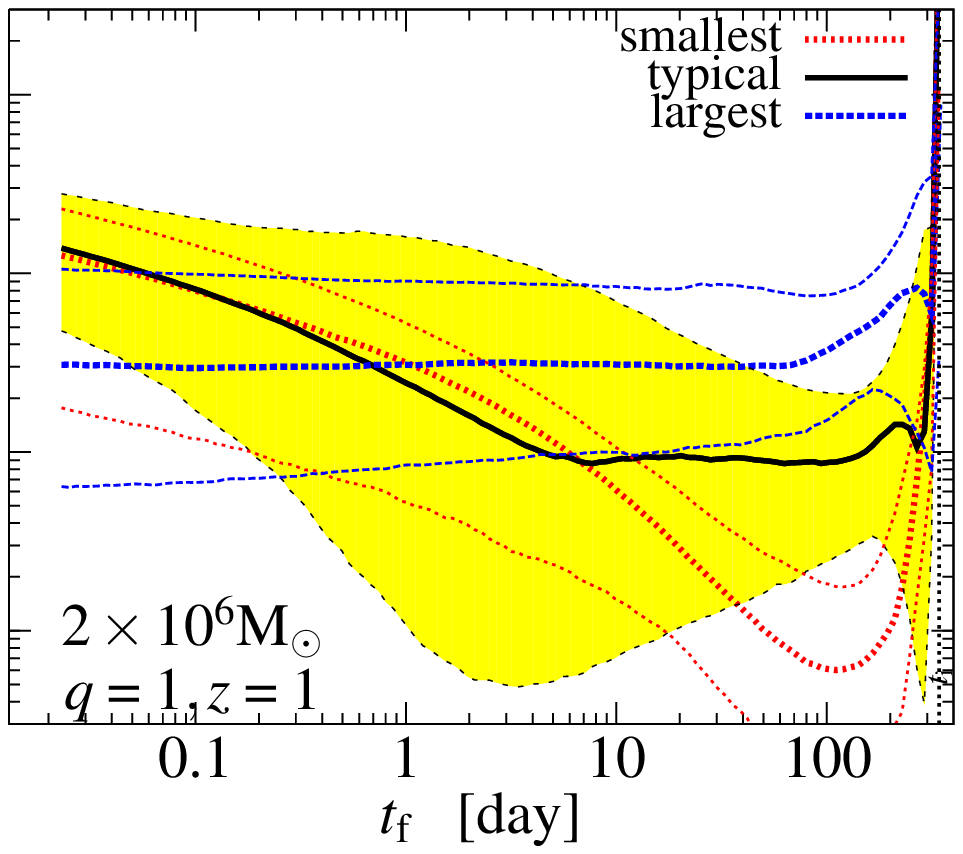}}
\mbox{\includegraphics[height=4.85cm]{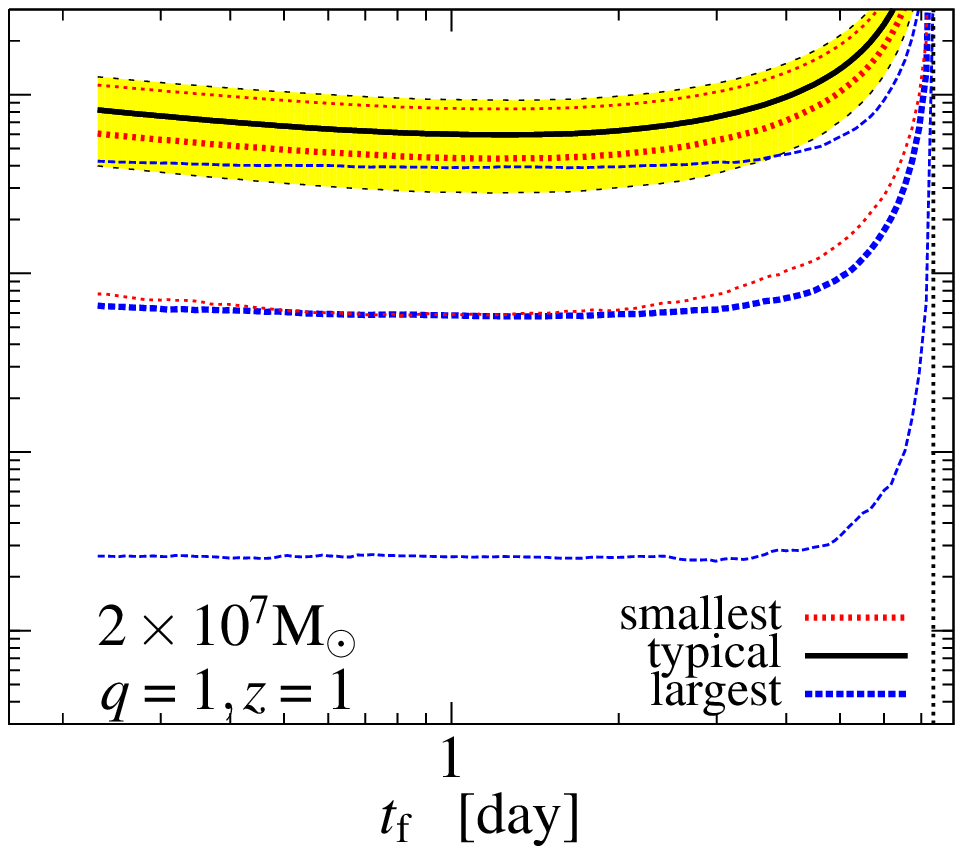}}
\caption{\label{f:ecc} Evolution with pre-ISCO look--back time, $\tf$,
  of the sky position error ellipsoid axis ratio ($a/b$) for the same
  six mass/redshift combinations (and equal masses $q=1$) as in
  Figure~\ref{f:errortau}. Each group of three curves corresponds to
  the $10\%$, $50\%$, and $90\%$ levels of eccentricity distributions,
  based on three bins (smallest, typical and largest) defined by the
  instantaneous value of the equivalent diameter $r$ at each $\tf$.
  Except for small observation times, the eccentricity remains small
  until the final day. This does not favor a strategy based on
  ``tiling'' the field with multiple pointings during a long-term
  monitoring campaign.}
\end{figure*}

Figure~\ref{f:ecc} shows the time--evolution of the axis ratio $a/b$
of the 2D angular error ellipsoids for the same mass/redshift
combinations as in Figures~\ref{f:errortau}--\ref{f:snapshot_b}.  To
look for possible correlations between size and eccentricity, we first
sorted all binaries according to the total area of the error ellipse
into three bins, containing the ``small'', ``medium'', and ``large''
ellipsoids, corresponding to 0--20\%, 40--60\%, and 80--100\% of all
events ranked by the instantaneous value of $r=\sqrt{ab}$.
Figure~\ref{f:ecc} then shows the evolution of the 10, 50, and 90
percentiles in eccentricity in each of these three bins
separately. The shaded region highlights the distribution of
medium--sized ellipsoids, and the thick curves correspond to the
median (50\%) eccentricities in each bin.  Depending on the binary
parameters, we find a systematic trend between the size and
eccentricity of the angular error ellipsoid.  Figure~\ref{f:ecc} shows
that if the ellipsoid is relatively small 3 months before merger, then
it is very circular, while the larger ellipsoids are more eccentric,
but subsequently, at a few days before ISCO, the smaller ellipsoids
become just as eccentric as the big ones.

Overall, Figure~\ref{f:ecc} shows that the eccentricity evolution of
the angular error ellipsoid is complicated, but that at the most
relevant epochs, i.e. in the last two weeks prior to merger, when the
absolute errors have shrunk to interesting levels, the eccentricity
increases but remains small until the final day. More specifically,
Figure \ref{f:ecc} shows that the eccentricity is significant in the
beginning, in the first few weeks after the binary enters {\it LISA}'s
frequency band.  However, the error ellipse then soon
``circularizes'', and the eccentricity remains small until the last
day before merger. The typical axis ratio decreases below 1.3, 1.1,
and 1.01 after two months of observation for the large, medium and
small ellipses, respectively. Except for the large ellipses, the
eccentricity tends to increase again in the last week before
merger. The axis ratio is typically under 1.3 before the last day to
merger, with a $10\%$ probability for $a/b=3$.

\subsubsection{Effects of Mass, Mass Ratio, Upper Harmonics, and Spins}
\label{sec:spins}

For high $M_z$ values, the signal frequency at ISCO is low, and the
binary coalesces soon after it enters {\it LISA}'s frequency band. For
example, the inspiral of a binary with $m_1=m_2=10^7~{\rm M_\odot}$
and $z\gsim 1$ can only be observed for less than 10 days. In such
cases, the sky localization eccentricity does not have time to
circularize and the eccentricity remains high (typically $a/b\sim
5$--$10$) throughout the entire observation. We note however, that
higher order GW harmonics beyond the restricted post-Newtonian
approximation become relevant in this case, since these harmonics are
inside the LISA frequency band much sooner. This improves angular
errors greatly for the high mass binaries \citep{Arun07b,ts07}. Due to
the increase in observation time, we expect higher GW harmonics to
circularize the angular ellipsoid for the more massive
binaries.\footnote{Note that the higher GW harmonics do not change the
general expectation that the localization errors can be deduced from
the slow amplitude modulation due to the detector motion.  Therefore,
compared to the restricted PN waveform, it is only the effective
increase in the observation time which results in a difference for the
localization ellipsoid.}

Although Figures~\ref{f:errortau}--\ref{f:ecc} correspond to equal
mass binaries, we also ran simulations for other mass ratios,
$q=m_1/m_2\sim 0.1$, and found very similar results. A plausible
explanation is that two competing effects approximately cancel each
other: (1) for fixed total mass $M$ but lower mass ratio $\eta$, the
instantaneous signal amplitude is lower ($h_{+,\times}(t)\propto
\eta$) while (2) the frequency evolves at a slower rate ($\D f/\D
t\propto \eta$ and the frequency at ISCO is independent of $\eta$)
implying that the binary spends more time at larger frequencies where
the detector is more sensitive. Relative to the equal mass case, these
competing effects approximately cancel out for $q=0.1$ for both the
size and eccentricity of the angular error ellipse. We find
differences only for the high mass systems where the total observation
time is much larger for $q=0.1$, leading to a smaller angular error
ellipsoid and lower eccentricity.

We expect this description to hold when including spin effects until
the final day prior to merger if the observation time is at least
months.  For high spin binaries, however, spin precession is expected
to add an important extra feature to the waveform during the last day
which improves the final measurement angular errors by a factor of
$\sim 3$ \citep{vec04,lh06}. In this case, we expect spin precession
effects to further increase the angular error eccentricity during the
last day. Indeed, recent results by \citet{lh07} show that the minor
axis acquires a long tail toward small values, $10''$--$1'$,
indicating that the best events at ISCO will possibly be very
elliptic.  On the other hand, including spin precession reduces 
the error eccentricity compared to our estimates if the total
observation time is less than a month (R. N. Lang, private
communication, 2007).  Therefore, spin precession effects will have a
tremendous impact on EM follow--up observations, allowing a
1--dimensional tiling of the field with small FOV instruments at late
times.

\subsubsection{Sky position dependence}

Figures~\ref{f:errortau} and \ref{f:ecc} show the basic statistics of
the expected localization errors when all six angular parameters (2
sky position, 2 angular momentum orientation, and 2 detector
orientation parameters) are chosen randomly.\footnote{Note that the
  measured signal depends on only 5 combinations of these angles,
  therefore only 5 variables are actually varied
  (\S~\ref{sec:assumptions}).}  This is useful to get an idea of the
range of possible accuracies we may have when the LISA data becomes
available.

It is also interesting to calculate how the LISA localization errors
depend on the actual sky position of the source, since EM counterpart
detections can have very different prospects, depending on this
position. For instance, it might be difficult to find a counterpart if
it lies in the Galactic plane, due to astrophysical foregrounds. On
the other hand, a specific example of a preferred direction could be
the Lockman hole, where the obscuring neutral hydrogen column density
is anomalously low.

To examine possible systematic effects with respect to the fiducial
sky position, we have divided the $(\theta_N,\phi_N)$ sky into cells
and calculated the distribution of errors for each cell by choosing
the remaining angular parameters randomly. Surprisingly, in a previous
work, \cite{mh02} showed that no significant trends exist at ISCO as a
function of the polar angle $\theta_N$.  Our computations (for
$m_1=m_2=10^6\Msun$ and $10^7\Msun$ at $z=1$) confirm this result to
the 10\% level, throughout the final 6 months of observation.  (Median
errors systematically get just slightly worse by less than 10\% for
sources that lie towards the Ecliptic.) Moreover, we find that
localization uncertainties are independent of $\phi_N$. This, however,
is not surprising since in Paper I we have shown that the LISA
measurement depends on the combination $\phi_N-\Phi$, and is otherwise
independent of $\phi_N$. Therefore the randomness of $\Phi_{\rm ISCO}$
perfectly smears out any systematic effects with respect to $\phi_N$.
We note that \citet{lh07} find a very different $\theta_N$ dependence:
they find Ecliptic equatorial sources to have $\sim 2$ times larger
angular uncertainties than sources near the poles at ISCO.  Although
\citet{lh07} employ a different set of assumptions (accounting for
correlations and spins, and choosing $\Phi=\Xi=0$ at the start of the
mission), these differences cannot be responsible for the strong
$\theta_N$--dependence.  The origin of this dependence therefore
remains unclear to us.

On the other hand, it is plausible to expect systematic effects in
both the Ecliptic latitude and longitude if measuring the longitude
relative to the detector position, $\phi_N-\Phi(t)$. We find this to
be a useful measure, since once the LISA plans are fixed, $\Phi(t)$ will
be known, and $\phi_N-\Phi(t)$ can be converted to Galactic
coordinates for each $\phi_N$ and $t$. For instance, depending on
$\Phi({\rm January})$, it might turn out that a ``good LISA
direction'' is in the Galactic plane in January and July, but not
during other months.

We calculate error distributions for cells of size $\Delta \cos
\theta_N \times \Delta (\phi_N-\Phi_{\rm ISCO}) = 0.1\times
10^{\circ}$ and find strong systematic effects: the median
localization error improves mostly in the instantaneous LISA plane
during the final month, and it is worst in the two orthogonal
directions $\pm{\bm N}_{\rm worst}(t)$, given by $(\theta_N,
\phi_N-\Phi(t))=(120^{\circ},0^{\circ})$ and
$(60^{\circ},180^{\circ})$.  This result has also been found
independently in a recent study by \citet{ts07} using non-restricted
PN waveforms for high mass binaries.  The angular resolution is worst
orthogonal to the LISA plane due to geometry: as our viewing angle of
the binary changes, due to LISA's orbital and rolling modulation, it
causes the smallest change in the waveform since it corresponds to an
extremum of the function [measured waveform] vs [viewing angle].
Depending on the source position relative to the plane of the
instantaneous orientation of the LISA triangle,
the median angular localization
uncertainty for $r(t)$, in the case $m_1=m_2=10^6\Msun$ and $z=1$,
ranges between 5'--50', 30'--60', 55'--90', 95'--135', and 170'--250'
at ISCO, 1 day, 1 week, 1 month, 3 months, and 6 months before ISCO,
respectively. Therefore, during the most relevant last week,
the area of the sky-position error ellipse can vary significantly,
by a factor of 4-100, depending on source position.
We find that within the last month, the median errors are roughly
rotation--symmetric around ${\bm N}_{\rm worst}(t)$.\footnote{Note
that this rotation symmetry might be a consequence of choosing
$\Xi_{\rm ISCO}$ randomly in our Monte Carlo simulation (see
\S~\ref{sec:assumptions}).  We also tried fixing $\Phi_{\rm
ISCO}-\Xi_{\rm ISCO}$ at some value and found an approximate rotation
symmetry in this case as well.}  Therefore, systematic effects become
prominent approaching ISCO because errors improve much more quickly
near the LISA plane.  Quite remarkably, the errors are small enough
very early on everywhere on the sky allowing to decide several months
in advance whether the merger will take place in the sensitive region
of LISA sky. {Therefore, if several coincident inspiral events are
identified during the LISA mission, it will be possible to choose
which one(s) to monitor with electromagnetic instruments according to
which event is projected to have the best LISA precision closer to the
merger}. (The broader issue of cutting down the number of potential
host galaxy counterparts will be discussed in
\S~\ref{sec:cutting} below.)

\section{Triggered Electromagnetic Observations}
\label{sec:monitoring}

The results summarized above suggest that it will be possible to
identify, prior to merger, a small enough region in the sky where any
prompt electromagnetic (EM) counterpart to a LISA inspiral event will
be located. Given sufficient ``advance notice,'' it will then be
possible to trigger a search for EM counterparts as the merger
proceeds (and also during the most energetic coalescence phase).

The several square--degree field will, of course, contain a very large
number of sources (a few $\times10^5$ galaxies in total, at the
limiting optical magnitude of $\sim 27$ mag that may be relevant; e.g.
\citealt{mad99} and discussion below).  Having predictions for the
spectrum and time--dependence of a coalescing SMBH binary, and
therefore knowing what to look for, would obviously greatly help in
identifying the counterpart, possibly allowing an identification even
post--merger.  Such a prediction, however, would require understanding
the complex hydrodynamics and radiative properties of circumbinary gas
at the relevant small separations of a few to a few thousand
Schwarzschild radii. This is a notoriously difficult problem even in
the much simpler case of steady accretion onto a single SMBH
\citep[e.g.][]{mefa01}.  This suggests that the best strategy may be
an ``open--minded'' search for any variable signatures prior and
during coalescence.

In the first half of this section, we describe several scenarios in
which one may expect an EM signal to be associated with the
coalescence of a SMBH binary (see also \citealt{dot06}).  Most of this
material in \S~\ref{subsec:expectedvariability} is a review of
existing literature. However, we also present some original material,
such as calculations of the relevant orbital time--scales
(Fig.~\ref{f:torbit}), a rough estimate for the spin magnitude and
orientation uncertainties for the binary prior to coalescence
(Eq.~\ref{e:spin}), and we propose that candidates may be selected in
advance based on whether or not the binary is expected to experience a
large and appropriately directed recoil that would favor bright EM
emission (\S~\ref{sec:transient}).

In most cases, reliable assessments of the nature of the potential EM
counterpart are difficult to obtain. What is most important for our
purposes is that these scenarios make plausibility arguments for
coalescing binaries being able to produce detectable variable signals
during their last few weeks prior to merger -- either with periodic
variations during the inspiral phase, or a strong transient signal
during the final plunge/ring--down/recoil phases. In the second half
of this section, \S~\ref{subsec:variabilitysearch}, we give
quantitative estimates of the capabilities of planned wide--field
instruments to look for such variable signals.

\subsection{Expected or Plausible Variable EM Signatures}
\label{subsec:expectedvariability}

\subsubsection{Periodic Flux Variations During the Inspiral Phase}

The problem of a SMBH binary embedded in an accretion disk has
received relatively little attention in the literature. It shares
similarities with that of a planet embedded in the protoplanetary disk
of its central star, which has been more extensively studied. The case
of two near-equal mass SMBH, embedded in a thin, co--rotating disk,
has been studied by a few authors.  Numerical simulations
\citep{al94,gunther04,escala04,escala05} indicate that, in this case,
a central hole is expected to open in the disk due to tidal effects
and the gas surface density within the binary's orbit could be
significantly reduced.  In the limit of efficient gap clearing, no EM
signal would be expected until after a viscous time ($t \gsim$ years)
past merger, when the inner edge of the gas disk falls in and starts
accreting onto the black hole remnant, producing an X-ray afterglow
\citep{mp05}.

However, in a more realistic situation, with no perfect azimuthal
symmetry, and especially for unequal masses or thick disks, some gas
may still flow across the binary's orbit, and ultimately accrete onto
one or both SMBHs.  It is not unreasonable to expect that this
inflowing gas could produce a bright and variable source, with
variability correlated with the orbital period.  Indeed, the Newtonian
potential around the binary will fluctuate periodically, with order
unity variations within a radius equal to the binary separation.  The
gas responds to this periodic perturbation, and a corresponding
quasi--periodic flux variability has been proposed on the basis of
recent numerical simulations \citep{macfadyen07,hayasaki07}.

Earlier numerical simulations, motivated largely to explain the 12--yr
variability of the blazar OJ 287, also found significant mass in--flux
across the edge of the central hole \citep{al96,gunther04}, especially
when the disk gets thick (although these models relate to $\sim 1000$
times larger separations than the GW-emitting stage of interest here).
Likewise, in the planetary context, close pre-main-sequence binary
stars are expected to clear central holes in their proto-planetary
disks, but \citet{jen07} argue that material can flow from the
circumbinary disk across the gap to explain the observed 19--day
variability of the young binary UZ Tau E.

There are other scenarios that may plausibly produce a variable
signature related to the binary orbit.  For example, \citet{am02}
examine the case of a highly unequal mass binary, and argue that a
significant amount of gas may be driven in by the merger if it is
initially present interior to the orbit.  Another way to promote
variable signatures associated with the influence of the binary's
orbital motion is to allow gas to flow in even as gravitational
radiation rapidly shrinks the binary's orbit. This may be possible if
the circumbinary disk is geometrically thick (or advection-dominated),
with a viscous timescale much shorter than traditional thin accretion
disks.

\begin{figure}[tbh]
\centering
\mbox{\includegraphics[width=8.5cm]{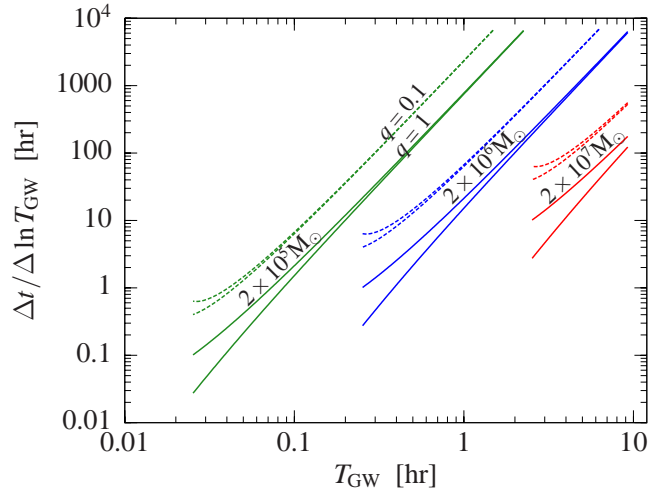}}
\caption{\label{f:torbit} The time spent (in hr) per logarithmic GW
  cycle-time bin during the inspiral for
  $M=2\times(10^5,10^6,10^7)\Msun$ (as labeled), $q=1$ (solid) and
  $0.1$ (dashed) for $z=1$ in the restricted 2.5PN approximation.  In
  each case two curves with the same style correspond to the two
  extremes: maximum aligned and anti-aligned spins; a general case lies
  in between these extremes.  $T_{\rm GW}$ is the GW cycle time in
  hours. The curves end at $\ti=1\yr$ or at $f_{\min}=3\times
  10^{-5}\Hz$, corresponding to $T_{\rm GW}=9.3\,\rm hr$.  Note that
  the number of GW cycles can be estimated as $N_{\rm cycles}=y/x$,
  from the $x$ and $y$--axis values corresponding to each curve.}
\end{figure}

Assuming that variability is indeed present, what is the expected
period of this variability? Again, the answer is uncertain, but a
reasonable guess is that the variability will be related to half the
orbital time--scale of the binary, corresponding to the periodic
quadrupolar perturbation of the gravitational potential.

In Figure~\ref{f:torbit} we show the time ($\Delta t$) spent by a
binary with a total mass of $10^{5,6,7}~{\rm M_\odot}$, redshift
$z=1$, at each orbital period $t_{\rm GW}$ (in logarithmic bins of
$t_{\rm GW}$).  The plots show cases with mass ratios of $q=1$ (solid)
and $0.1$ (dashed).  The various curves of the same type show the
upper and lower bound of the time-evolution depending on the magnitude
and orientation of BH spins. The presence of spin modifies appreciably
only the final day of the merger.

Quite remarkably, Figure~\ref{f:torbit} shows that the expected EM
variability timescale is in a very fortunate range for detection:
between several minutes up to 10 hours within the last year of merger.
Detecting slower variations would require several exposures over
exceedingly long time--periods, while detecting faster variations (of
the same amplitude) would require much higher sensitivity per unit
exposure time. If a periodic electromagnetic variation is found within
this range, e.g. for an AGN in X-rays, it would indicate evidence for
an SMBH binary approaching coalescence.  The evidence would be
especially convincing if this nearly periodic variable
``light--curve'' tracked the known, nearly periodic GW emission, with
a significant cross--correlation between the periods of the two
waveforms.  The EM lightcurve may perhaps even follow the slow
decrease in period, related to the GW chirp, shown in
Figure~\ref{f:torbit}.

On the basis of the relatively well--defined nature of these potential
counterparts, it may even be interesting to carry out, well before the
LISA mission becomes operational, systematic searches for such
quasi-periodic behavior with the various all sky variability surveys
currently being planned.

\subsubsection{Transient Signatures During Final Coalescence}
\label{sec:transient}

Perhaps the most promising phase for the emergence of bright EM
counterparts to SMBH merger events is, in fact, during or shortly
after the violent final coalescence stage. This is the most energetic
phase of the binary's evolution, with typically $\sim 10^{58}$~ergs of
gravitational radiation released to the environment in a short $\sim
10^3$~s. There are at least two ways to couple this energetic outburst
to the immediate gaseous environment of the SMBH binary (see also
\citealt{mp05}).

A few percent of the binary system's rest mass is lost when the
binding energy of the two holes is carried away by GWs. Most of this
loss occurs during the final cycles of inspiral, the plunge and
subsequent ring-down. Gas bound to the SMBH's binary would thus
experience a sudden reduction in the mass of the central object, on a
timescale typically much shorter than the longer (local) dynamical
timescale on which this gas can respond. Although this sudden
disturbance is of moderate amplitude ($\sim$ a few $\%$), it is a
potential source of transient variability associated with the
coalescence, depending on details of the disturbed surrounding gas
properties \citep{boph07}.

A second, and possibly more direct and promising avenue to couple the
energetic GW burst at coalescence to the SMBH binary's environment is
through the GW recoil/kick experienced by the system. The accumulation
of linear momentum leading to the final recoil builds during the final
cycles of inspiral, the plunge and subsequent ring-down, so that it
appears as quasi-instantaneous except for any gas that may be located
in the immediate vicinity of the coalescing binary. The magnitude of
the kick is determined by the mass ratio, the magnitude of the spins,
and the relative orientation of the spin and orbital angular momentum
vectors.  The recoil kicks can be calculated accurately using
numerical relativity \citep[e.g,][]{bak06}. The final BH velocity is
found to reach $170 {\,\rm km}/{\rm s}$ for no spins \citep{gon07},
and $2500$--$4000~{\rm km}/{\rm s}$ for high spins in special
configurations \citep{cam07,bru07}.  For comparable mass binaries with
large spins and arbitrary orientation, a recoil $>500 {\,\rm km}/{\rm
  s}$ has a $30\%$ probability \citep{sb07}. If the spin axis is
aligned with the angular momentum, as maybe the case in gas rich
mergers, then the recoil is expected to be $\lsim 200 {\,\rm km}/{\rm
  s}$ \citep[and see \S~\ref{sec:cutting} below]{brm07}.  Even for a
modest recoil velocity of $100{\,\rm km}/{\rm s}$, this amounts to
$\sim 10^{53}$~ergs of mechanical energy that can in principle be
deposited in the environment as a SMBH remnant of $\sim
10^6$~M$_\odot$ ploughs through.  The EM signatures that may emerge
from such interactions are difficult to evaluate {\it a priori} since
they will depend on many details, such as the properties of the gas
that remains bound to the recoiling SMBH, the geometry of the recoil
system with respect to its environment, and possibly the remnant BH
spin.  We will simply note here that these EM counterparts are not
necessarily bound by the Eddington limit traditionally applied to
steady systems, if the recoil energy is deposited fast and efficiently
enough for the event to adopt an outflowing/explosive character.

For the present discussion, it is interesting to ask whether the final
recoil magnitude and direction can be estimated in advance, and to
determine whether it lies in the plane of the binary, where a gaseous
disk may be preferentially located. If so, then this advance knowledge
could possibly be used to pick the events that are most likely to
produce a transient EM counterpart. Since recoil velocities are
determined by the mass ratio, spin magnitudes and orientations, the
question is whether these parameters can be estimated with sufficient
precision in advance. Without specific computations, we can estimate
the advance uncertainties for these parameters using simple scaling
arguments. The mass ratio is determined from the 1PN effect, hence it
is a ``fast parameter'' that will likely improve quickly with time,
close to $(S/N)^{-1}$ (see \S~\ref{sec:assumptions}).  Spin effects
arise at 1.5PN order, implying that the binary spins are also ``fast
parameters''.  The final errors on the spin magnitudes are excellent
(see \citealt{lh06}), better than 1\%--6\%--20\% for $z=1$--3--5 if
$m_1=m_2=10^6 \Msun$. Even the final spin angle uncertainties have not
been estimated previously, but we can assume that they are not much
worse for large spins than the angular momentum orientation
uncertainties $(\delta \theta_L,\delta \phi_L)$, since the two are
coupled through precession, and precession was shown to help in
reducing the $(\delta \theta_L,\delta \phi_L)$ errors.  If the spin
magnitude is chosen randomly, then the spin orientation measurement
error is $3.5\deg$ for $m_1=m_2=10^6 \Msun$ at $z=1$ (R. N. Lang,
private communication, 2007).  We assume conservatively that the
evolution of spin errors scale with $(S/N)^{-1}$ until the last day
before ISCO.  According to Paper I, $S/N \propto \tf^{1/3}$.  In this
case
\begin{equation}\label{e:spin}
        \delta {\rm spin}(t_f) \simeq \left(\frac{t_f}{t_{\rm
        ISCO}}\right)^{1/3}\delta {\rm spin}(t_{\rm ISCO}),
\end{equation}
with $t_{\rm ISCO}=33 \min \times (M_z/(4 \times 10^6 \Msun))$. For
$m_1=m_2=10^6 \Msun$ at $z=1$, we find $\delta {\rm spin} \simeq 1\%
\times [t_f/(33 \min)]^{1/3}$.  i.e. spin magnitudes will be
determined to $\sim 8\%$ at 10 day before ISCO. For the direction,
$\sim 10$~deg ($\sim 20$~deg) precision can be achieved 10 days before
ISCO, if the final uncertainty at ISCO is 1.3~deg (2.6~deg).
This information could thus in principle be used to focus on events
with specific magnitudes and/or directions for the expected recoil, in
an effort to maximize the likelihood of finding bright EM counterparts.

\subsection{Searching for Variable Signals with Wide Field Instruments}
\label{subsec:variabilitysearch}

\subsubsection{Basic Parameters for a Variability Search}

The most obvious observational strategy to catch a prompt EM
counterpart, especially in the absence of robust theoretical
predictions, is to blindly and continuously monitor the source area,
through coalescence, for variable sources.  The requirements for such
a monitoring -- large FOV, and a fast camera -- coincide with those of
searches for distant Supernovae (SNe) in the optical bands.  Motivated
by the latter, there are several instruments being built or planned
with specifications that could prove suitable for a {\it LISA}
counterpart search.  Table~\ref{tab:surveys} shows a non-exhaustive
list of facilities, either planned or being built, that have a several
square--degree field of view and may be suitable for a {\it LISA}
counterpart search.

\begin{deluxetable*}{lrrrrl} \tablecolumns{6}
\tablewidth{0pt} \tablecaption{\label{tab:surveys} Wide Field Electromagnetic Instruments} \tablehead{\colhead{Name} & \colhead{FOV $\left({\rm deg}^2\right)$}
& \colhead{bands} & \colhead{mag. limit (exp)} & \colhead{ start date} & \colhead{reference}} \startdata
High Energy Transient Explorer-2  & 0.9, 1.6 srad     & soft, hard X-ray, $\gamma$ 
 & $8\times 10^{-9}\frac{\rm erg}{{\rm cm}^2{\rm s}}$ & 2000 & space.mit.edu/HETE \\
Galaxy Evolution Explorer & 1.2 & Near, Far UV 
                                        & 21-22 (1 hr) & 2003 & www.galex.caltech.edu \\
Palomar--Quest      				& $3.6\times4.6$ & $U,B,R,I,r,i,z$ & 17.5-21.0 (140sec)      & 2003 & \citet{bal07} \\
XBo\"otes Survey 		& $9.3$ & X-ray & $(1$--$8)\times 10^{-15}\frac{\rm erg}{{\rm cm}^2{\rm s}}$ & 2005 & www.noao.edu/noao/noaodeep\\&&&&&~~/XBootesPublic \\
SkyMapper  			& $2.3\times2.4$ & $u,v,g,r,i,z$   & 20.6-21.9 (110sec)      & 2007 & www.mso.anu.edu.au/skymapper \\
South Pole Submillimeter Telescope   &  $1$  & sub-mm (350--850)$\mu$m & 1 mJy (18-27 hr) & 2007 & cfa-www.harvard.edu/$\sim$aas/tenmeter \\
Low Frequency Array     &  $20$ & radio (10--240)MHz  & (0.03--2)mJy (1 hr)  & 2007 & www.lofar.org\\
\hline
Pan-STARRS-1 & $2.5\times2.5$ & $g,r,i,z$       & 24 (1 min, 5$\sigma$)  & 2008 & pan-starrs.ifa.hawaii.edu \\
Gamma-ray Large Area Space Tel. & $ 2\,\rm srad$ & $\gamma$ 
					& $6\times 10^{-9}\frac{1}{{\rm cm}^2{\rm s}} $ (1yr)  & 2008 & glast.gsfc.nasa.gov \\
Dark Energy Survey & $2.2\times2.2$ & $g,r,i,z$ & 24 (100 min,$5\sigma$)       & 2009 & www.darkenergysurvey.org \\
Kepler             & $105$ & 430--890nm & 24 (150 min,$5\sigma$) & 2009 & kepler.nasa.gov \\
Large Synoptic Survey Telescope      &  $10$ & $u,g,r,i,z,Y$   & 27 (1 hr, 5$\sigma$) & 2014 & www.lsst.org \\
Joint Dark Energy Mission   & $15$ & 6 optical + 3 NIR & 27 (1 hr)       & 2016 & universe.nasa.gov/program\\&&&&&~~/probes/jdem.html \\
Wide Field Imager & $1$ & 350--1800 nm & 24 (110 min) & 2017 & sci.esa.int/science-e/www/object\\&&&&&~~/index.cfm?fobjectid=39692 \\
Energetic X-ray Imaging Survey Tel. & 154 x 65 & X-ray & $2 \times 10^{-14}\frac{\rm erg}{{\rm cm}^2{\rm s}}$ & 2020 & exist.gsfc.nasa.gov\\
Square Kilometer Array      & $1$    & radio (20--200)GHz  &  1 $\mu$Jy (1 hr)        & 2020 & www.skatelescope.org
\enddata
\tablecomments{The table shows EM instruments, existing, being built,
or planned, that have a several square--degree field of view and may
be suitable for a {\it LISA} counterpart search.  The 3rd column shows
the observational wavelength bands, and the 4th column shows the
sensitivity that can be reached in these bands in the indicated
integration time at the quoted S/N (derived from information at the
websites listed for each instrument in the 6th column, when possible).
}
\end{deluxetable*}

According to the arguments in the previous section, a coalescing
binary may produce some variable signature in the last two weeks
before ISCO. The amplitude and period of any variability, however is
highly uncertain.  In view of this uncertainty, it is interesting to
work backwards, and compute the expected brightness of a variable
signal that is some fixed fraction of the Eddington luminosity.  An
Eddington-luminous SMBH at $z=1-3$, with a mass of
$(10^6$--$10^7)\,{\rm M_\odot}$, assuming a 10\% bolometric
correction, would appear as a $\sim 22$--25~mag point source in the
optical bands.  This implies, for example, that if the variable signal
is $\sim 10\%$ of the Eddington luminosity, then we need sensitivities
corresponding to 24--27~mag.  Ten times lower mass BHs would require
sensitivities of 26--29 mag. This variability is to be detected on an
orbital time--scale, which is between several minutes to several hours
(Fig.~\ref{f:torbit}).

Scaling from the sensitivities shown in Table~\ref{tab:surveys},
assuming $S/N\propto \sqrt{t}$, we see that a sensitivity of $\sim
25-27$~mag can be reached in $\sim 1$min to $\sim 1$hr integrations
with Pan--STARRS and with LSST.  We conclude that an LSST--class
instrument, repeatedly integrating over the {\it LISA} area, would be
able to detect a variability that corresponds to $\sim 10\%$ of the
Eddington luminosity and follows the period of the binary's orbit.

Figures~\ref{f:errortau} and \ref{f:ecc} show that the eccentricity of
the {\it LISA} error box may be significant during the last day or for
high--mass SMBHs with $M\gsim 2\times 10^7\Msun$, with the more
uncertain (major) axis longer by up to a factor of several than the
better determined (minor) axis.  This may impact the plausible
duration of any triggered monitoring.  If the telescope is rapid
enough, then one requires only that the minor axis fits within the FOV
(assuming that the FOV is spherical), since it will be possible to
tile the rest of the {\it LISA} ellipse by a few extra pointed
observations, displaced along the major axis.  On the other hand, if
the required integration time is comparable to the expected orbital
period, then performing such a tiling would mean shallow observations
that may miss the variability.  In this case, the monitoring campaign
can start only after the major axis has shrunk below the linear size
of the FOV.

\begin{figure}[tbh]
\centering{
\mbox{\includegraphics[width=8.5cm]{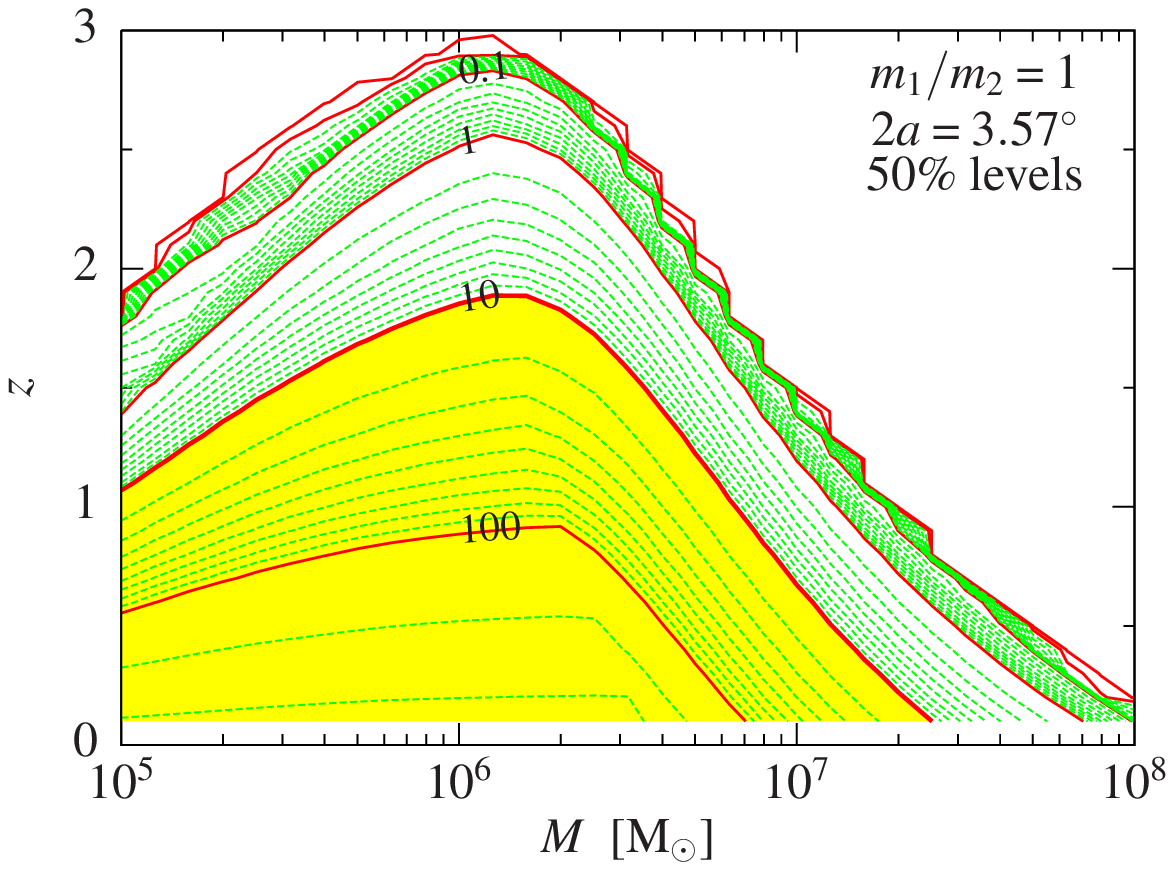}}\\
\bigskip
\mbox{\includegraphics[width=8.5cm]{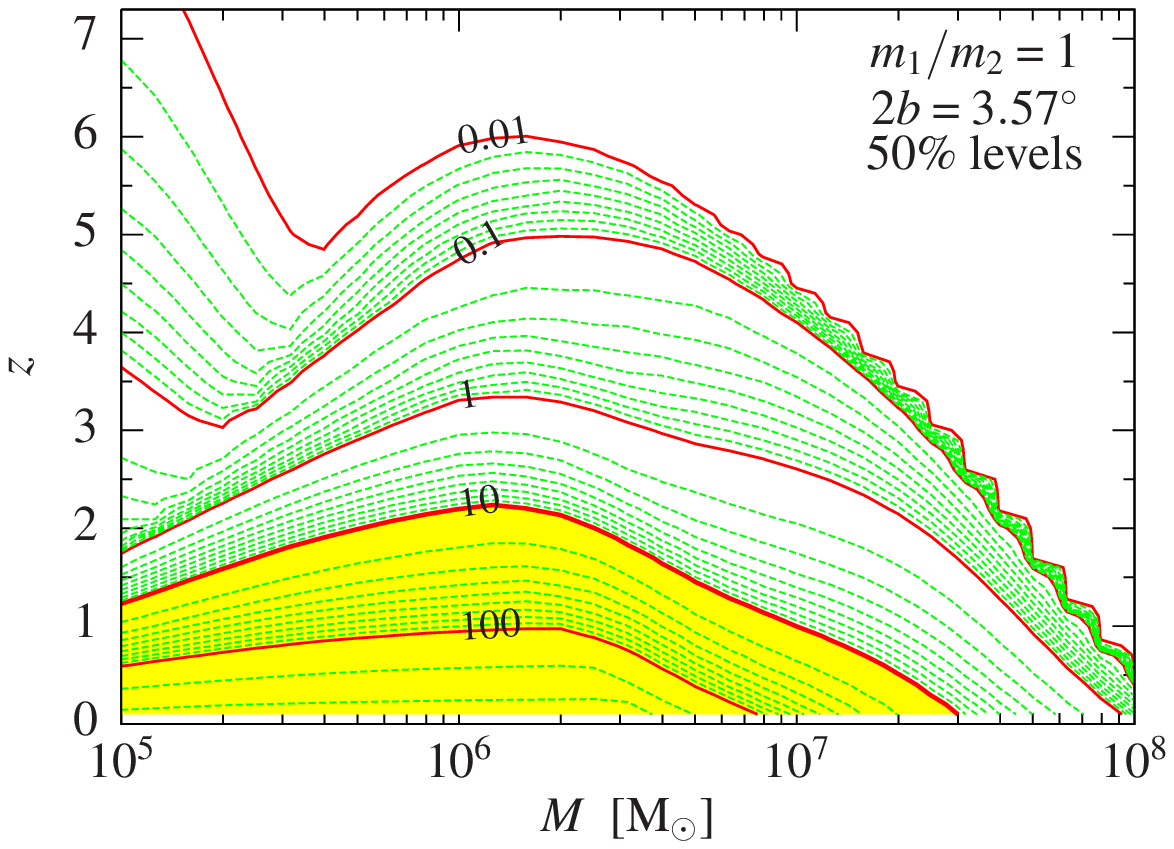}}}
\caption{\label{f:warning50} Contours of advance warning times in
the total mass ($M$) and redshift ($z$) plane, for typical events
($50\%$ level of cumulative distributions for random orientation
events) with SMBH mass ratio $m_1/m_2=1$.  The contours trace the
look--back times at which the major axis ($2a$, top panel) and the minor
axis ($2b$, bottom panel) of the localization error ellipsoid first
reach an LSST-equivalent field-of-view ($3.57^{\circ}$).  The contours
are logarithmically spaced in days and $10$ days is highlighted with a
thick curve. }
\end{figure}

\begin{figure}[tbh]
\centering{
\mbox{\includegraphics[width=8.5cm]{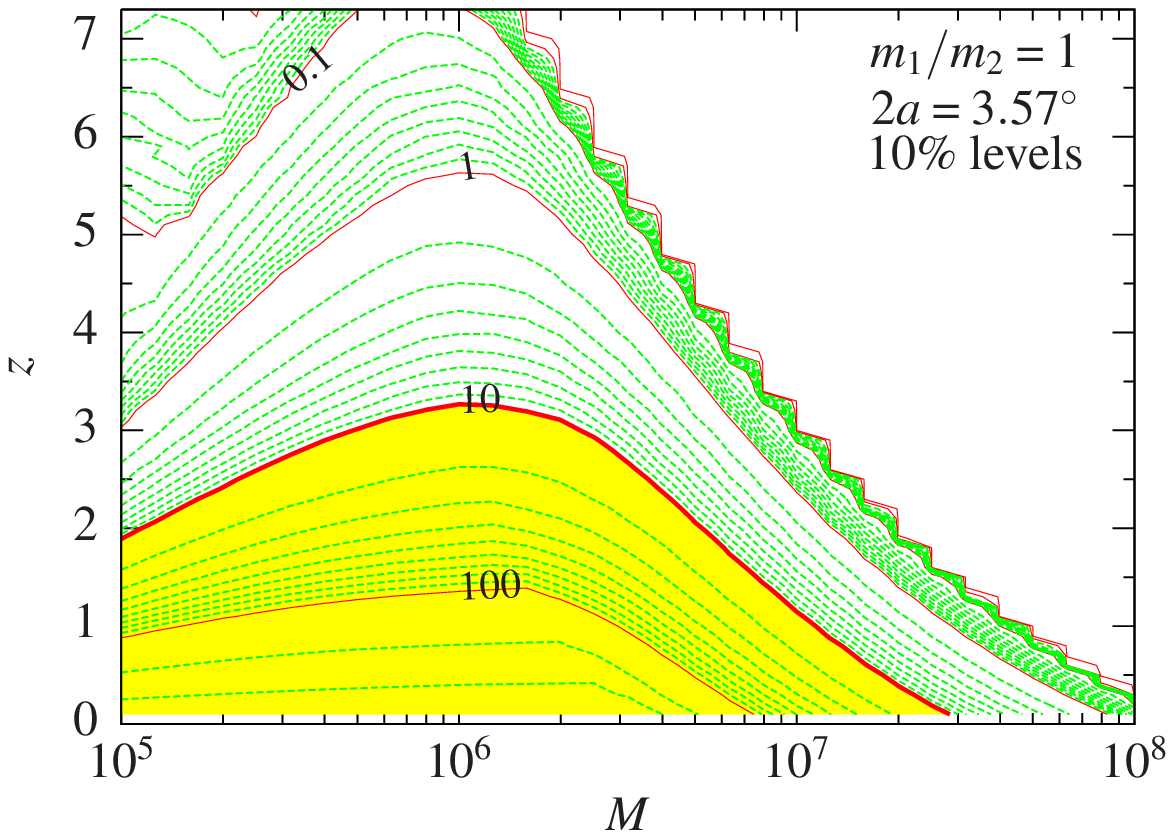}}\\
\bigskip
\mbox{\includegraphics[width=8.5cm]{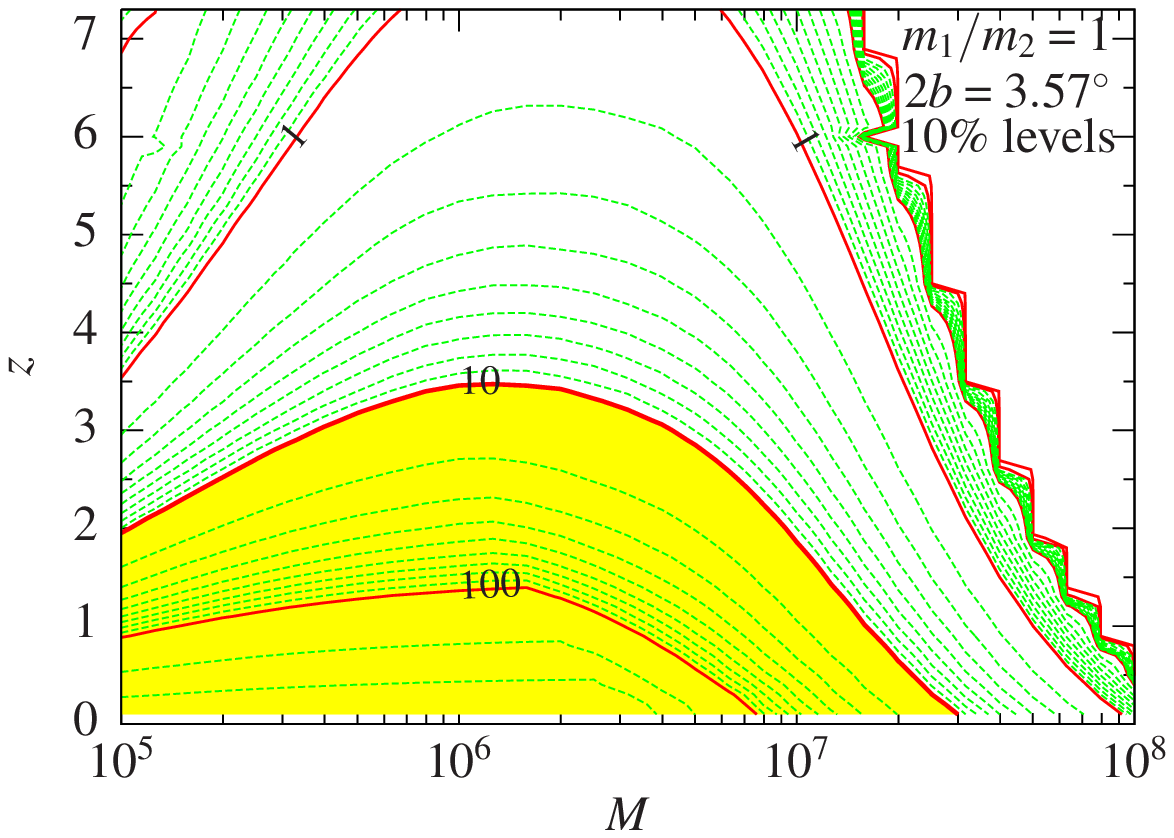}}}
 \caption{\label{f:warning10} Same as Figure~\ref{f:warning50}, except
 that results for the best ($10\%$ cumulative level) events are
 shown.}
\end{figure}

Figures~\ref{f:warning50} and \ref{f:warning10} display advance
warning time contours for typical ($50\%$) and best case ($10\%$)
events, adopting the LSST FOV as a reference.  Advance warning time
contours are logarithmically spaced, with solid contours every decade
and the shaded region highlights the $(M,z)$ region where at least
$10$--day advance notice will be available. The top panel of
Figure~\ref{f:warning50} shows that $10$ day advance warning to cover
the full error ellipsoid with a single LSST pointing is possible for a
large range of masses and source redshifts, up to $M\sim 3\times
10^7\Msun$ and $z\sim 1.7$. The bottom panel shows that the $(M,z)$
region in which this same $10$ day warning is limited to the minor
axis of the error ellipsoid reaches to somewhat higher
redshifts. Figure~\ref{f:warning10} shows how far the advance warning
concept can be stretched, by focusing on the $10\%$ best cases of
random orientation events. The top panel shows that a 10 day warning
to cover the full error ellipsoid with a single LSST-type pointing is
possible up to $z\sim 3$ for masses around $M \sim 10^6 \Msun$.  The
bottom panel shows that the $(M,z)$ region in which the same 10 day
warning is restricted to just the minor axis of the error ellipsoid
reaches out to comparable redshifts, for a similar range of masses.

As explained above, a unique variable EM signal could be
produced in the last stages of the merger.
Figures~\ref{f:warning50} and Figure~\ref{f:warning10} show that
requiring a warning of just one day would extend considerably the
range of masses and redshifts for which a single LSST-type pointing
is sufficient, out to $z\gsim 5$ for the best events.  However, as
we mentioned above, \citet{lh07} find that in the last day,
parameter correlations can degrade the errors by a factor of 2--3,
but this degradation is more than offset by the factor of several
improvement in the errors caused by spin precession (for high spin).
Therefore, our localization errors for the last day are overly
optimistic, by a factor of 2-3, for low spin mergers, but are
conservative for high--spin mergers.  Also, the issue of
eccentricity turns out to be most important in this regime (i.e. for
localizing the best events $\sim 1\,$day prior to ISCO, especially
for high spins, see \S~\ref{sec:spins}).  In this case, requiring
the FOV to cover only the minor axis would allow going out to $z\sim 8$. 

It should be possible to significantly improve on the above magnitude
requirements, if the signal is indeed close to periodic at $t_{\rm
GW}$.  This is because the above calculation requires the flux
variations to be detectable for any individual source in the LSST FOV.
A lower--amplitude variable signal would be lost in the noise for any
individual source (i.e. it will be consistent with photometric noise).
Fitting to a sinusoid--like template would require much less
sensitivity.  Even in the absence of a precise template for the EM
emission, one could use the GW waveform itself (or variations of
it). Cross--correlating the EM flux with the peaks and troughs of the
GW waveform should allow one to recover such below--the--threshold
variability, effectively decreasing the smallest detectable
variability amplitude by a factor of $\sim \sqrt{N_{\rm cycles}}$,
where $N_{\rm cycles}\approx 10^{3,2,1}$ is the number of orbital
cycles for the binary during the last two weeks, for
$M=10^{5,6,7}\Msun$, respectively (see
Fig.~\ref{f:torbit}).\footnote{Note that $\sqrt{N_{\rm cycles}}$ scaling is
correct only for constant modulation power per cycle, which might not
be the case for the EM variations.} This type of template matching
could presumably be attempted in various regions of the EM spectrum
(see Table~1 for a few possibilities).

\subsubsection{Cutting Down on the Counterpart Candidates}
\label{sec:cutting}

The above scenario so far envisions monitoring all sources in a few
square degrees, in $\sim$ hourly intervals, for $\sim$ two weeks prior
to and around ISCO.  This wide field will contain a very large number
of galaxies.  As seen earlier, the interesting optical magnitude limit
may be $\sim 25-27$, for variability at the level of $1-10\%$ of the
Eddington limit. The surface density of galaxies at this magnitude is
known from deep optical observations to be about $\sim 10^6\, {\rm
galaxies}/(10\, {\rm deg^2})$ \citep[e.g.][]{mad99}.  While this large
number of possible candidates may seem daunting, there are several
ways to significantly shrink the candidate list.  We enumerate these
options in this section.

{\em Final Angular Localization Cut.}  In Figures~\ref{f:warning50}
and \ref{f:warning10} we showed the advance warning times when the
source localization can be first covered with a $10\deg^2$ instrument.
However not all galaxies are going to be relevant in this region
eventually, since the area of the error ellipsoid shrinks continuously
(Fig.~\ref{f:errortau}).  For $m_1=m_2=10^6\Msun$ at $z=1$, for
instance, the $10\deg^2$ localization is typically reached at
$\tf\approx 86$ days.  The $(10$--$90)\%$ distribution levels of the
sky position uncertainty later decrease to $\delta\Omega = (0.1$--$1)$
and (0.6--7)${\,\rm deg}^2$ at $\tf=1$ and $10$ days before ISCO
(Fig.~\ref{f:errortau}). During the last day, the size of the
ellipsoid decreases by another factor of $\sim 3$ for a source with
large spins \citep{lh07}.  Therefore the number of candidates among
the galaxies within the original $10\,\deg^2$ field decreases to a
fraction $(6$--$70)\%$, $(1$--$10)\%$, and $(0.3$--$10)\%$, at 10
days, and 1 day before merger, and at ISCO, respectively. Obviously,
these cuts can only be applied progressively as the merger progresses,
always abandoning candidates that move out of the shrinking error
ellipse, and not immediately at the beginning of the monitoring
campaign.

{\em Photometric Redshift Slices.}  All of the survey instruments
listed in Table 1 should be able to observe the field in several
bands, and, for example, LSST should be able deliver photometric
redshifts of all sources with an accuracy of $\delta z\lsim 0.05$
\citep[see, e.g.][]{photoz}.  If LSST has already surveyed the target
area before the {\it LISA} event, then galaxies will have tabulated
magnitudes and photo--$z$'s already. These data may, however, not
pre--exist, or may not go deep enough (if the SMBH system has too
small a mass).  But the cumulative exposures of this particular
$\sim$~deg$^2$ field during a monitoring campaign should be sufficient
to reach target errors already in approximately the first day of
monitoring.

\begin{figure*}[tbh]
\mbox{\includegraphics[height=4.85cm]{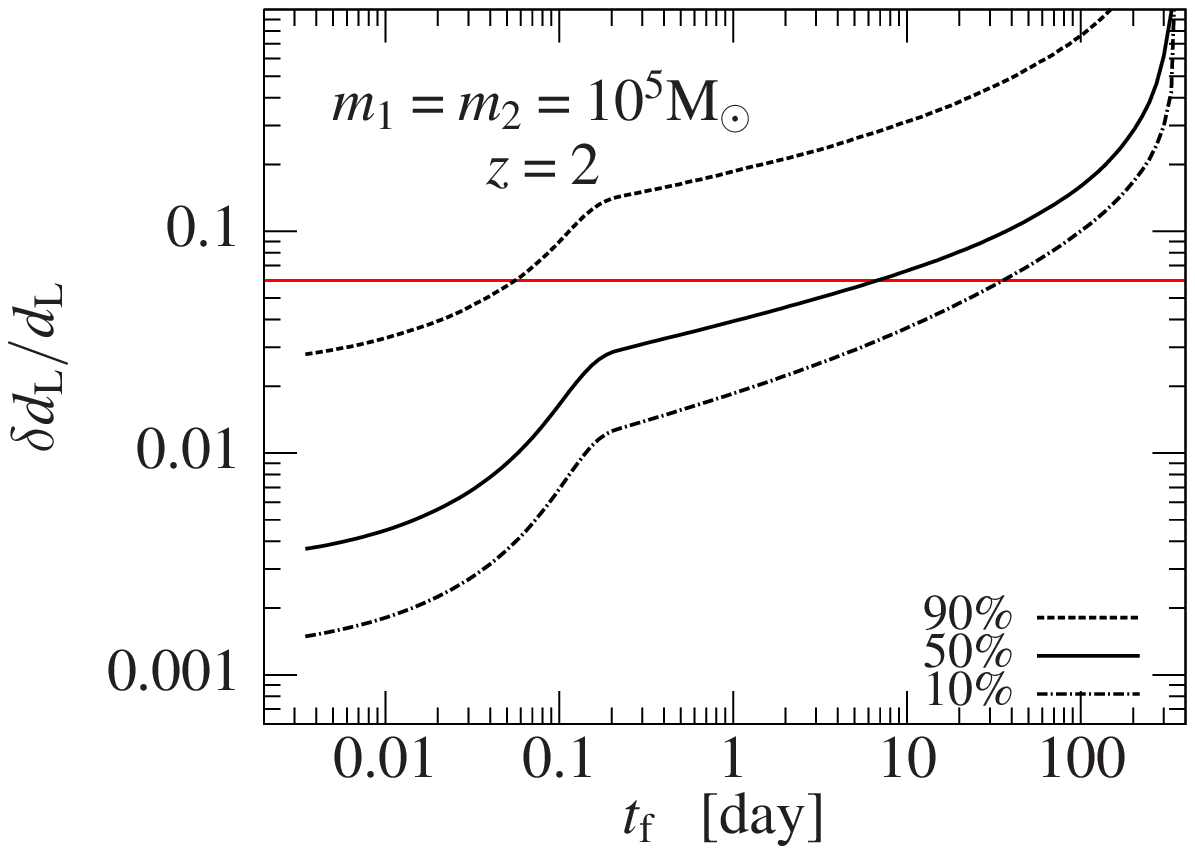}}
\mbox{\includegraphics[height=4.85cm]{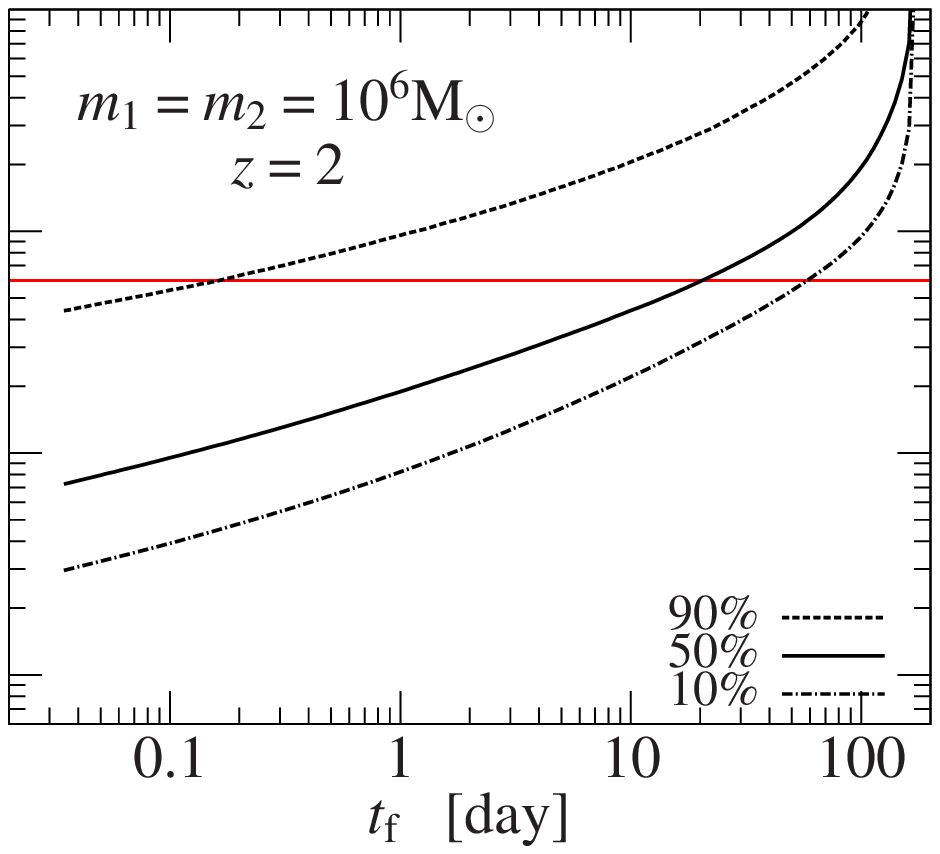}}
\mbox{\includegraphics[height=4.85cm]{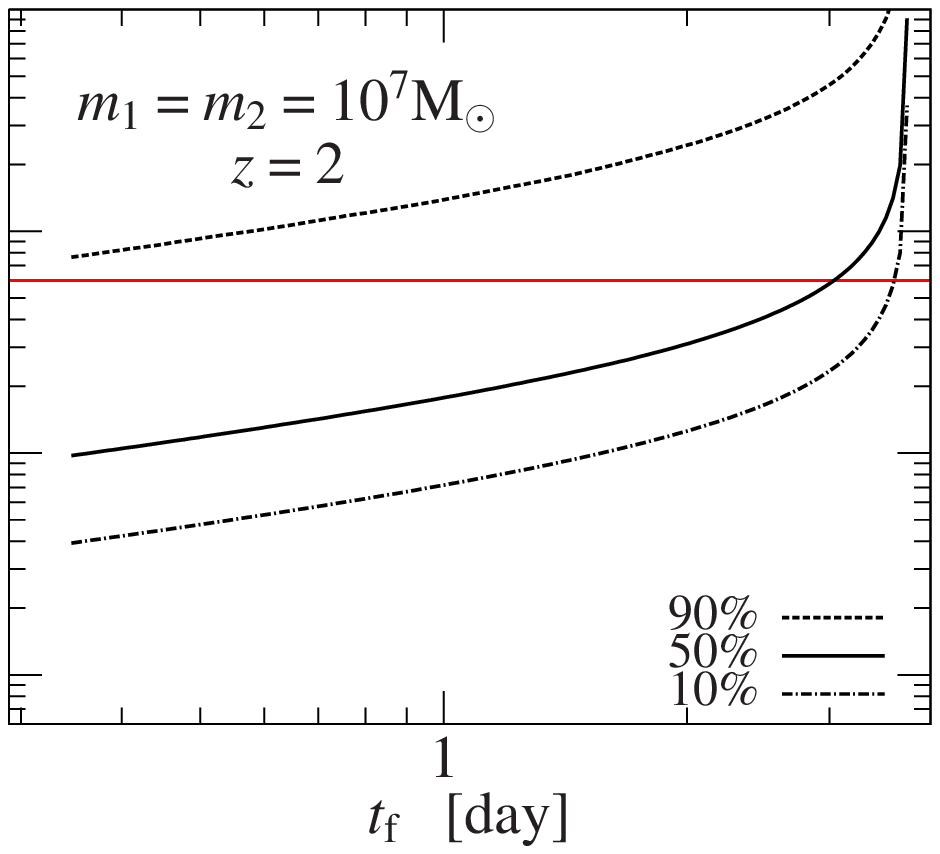}}\\
\mbox{\includegraphics[height=4.85cm]{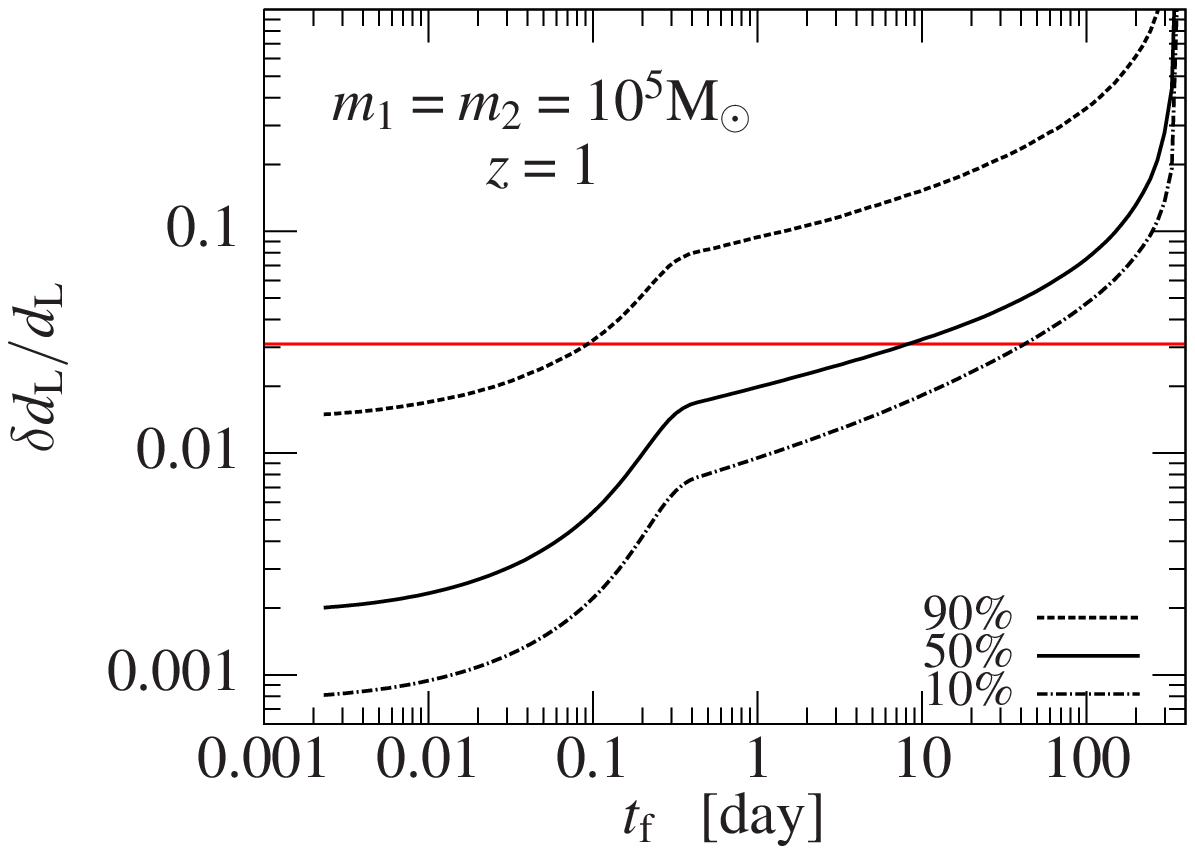}}
\mbox{\includegraphics[height=4.85cm]{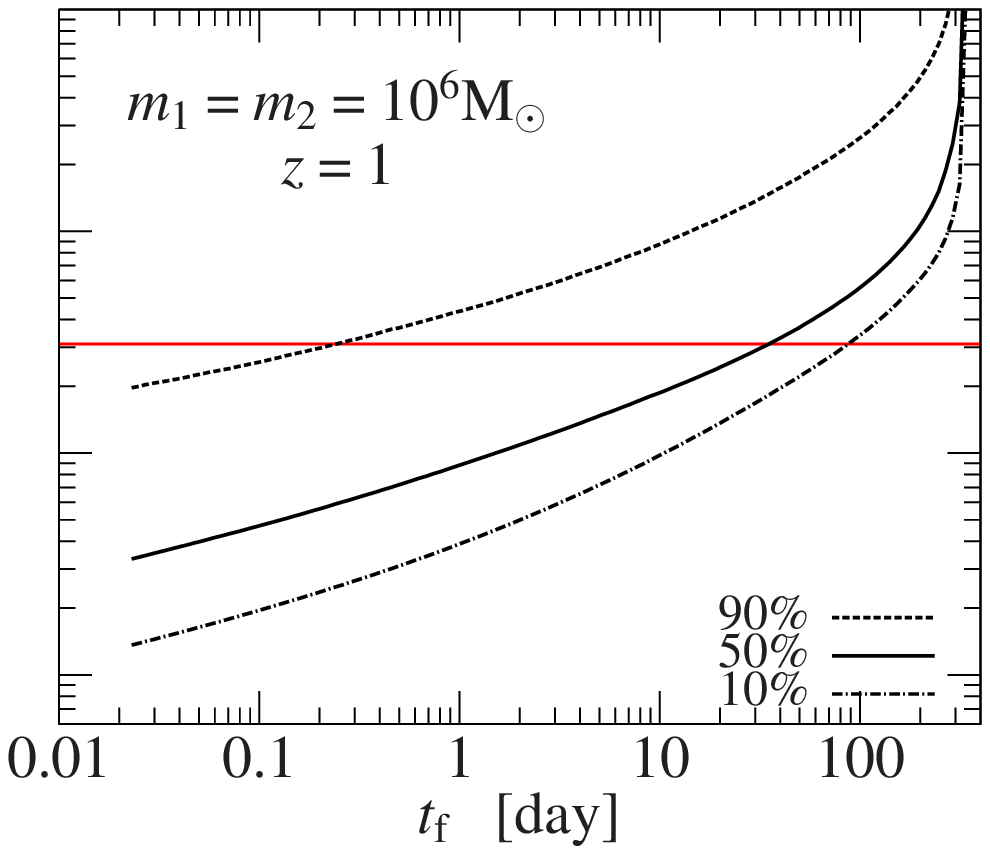}}
\mbox{\includegraphics[height=4.85cm]{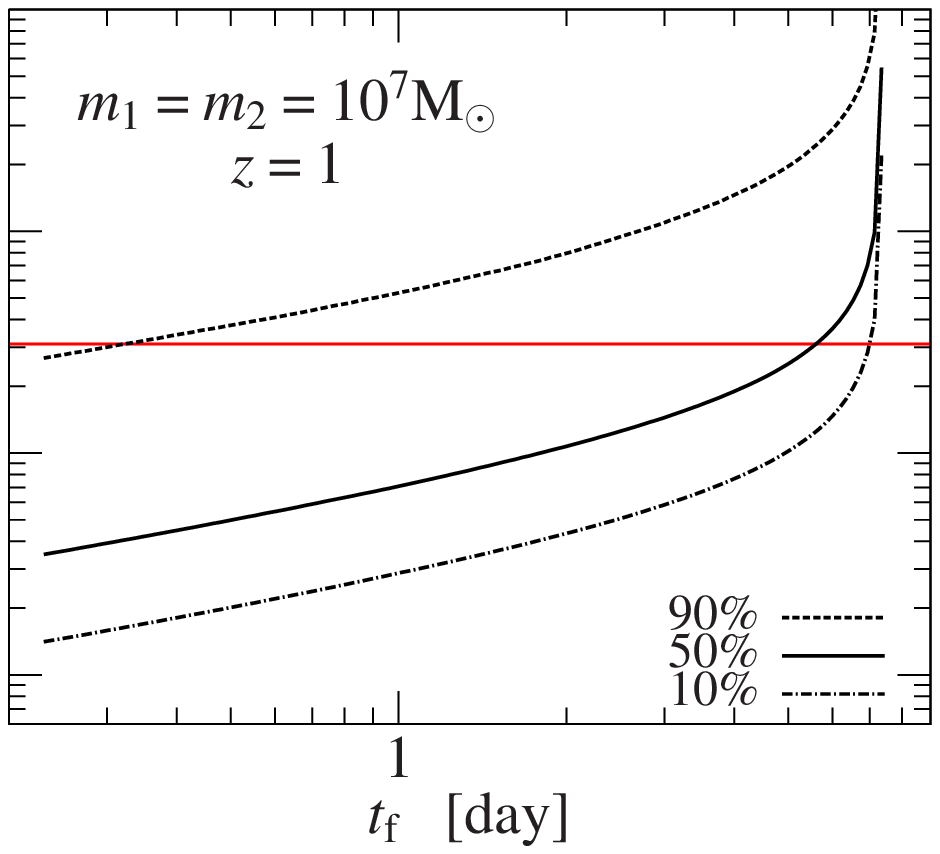}}
\caption{\label{f:dl} Evolution with pre-ISCO look--back time, $\tf$,
  of LISA errors on the luminosity distance $d_L(z)$ for
  $m_1=m_2=(10^5,10^6,10^7)~\Msun$ and $z=(1,2)$ as labeled.  The
  $10\%$, $50\%$, and $90\%$ levels of cumulative error distributions
  for random orientation events are shown. Horizontal lines delineate
  the level of uncorrected weak lensing uncertainties.  }
\end{figure*}

Importantly, {\it LISA} will deliver the luminosity distance $d_L(z)$
with high accuracy. In Figure~\ref{f:dl}, we show the errors on
distance as a function of time.  For a flat $\Lambda$CDM cosmology
with $(\Omega_{\Lambda}, \Omega_{m}, h)=(0.7,0.3,0.65)$ the redshift
uncertainty is given by $\delta z = (0.81,1.6) \delta \dL/\dL$ for
$z=(1, 2)$, respectively, provided that $\delta z>0.005$
\citep{koc06}.  Few percent errors are thus reached several days prior
to ISCO, except for the worst--case events, in the entire mass range,
$10^5-10^7~{\rm M_\odot}$, out to $z= 2$.  However, the distance
determination is ultimately limited by weak lensing uncertainties due
to fluctuations in the dark matter density along the line of sight,
shown as a horizontal line in Fig~\ref{f:dl} at $\delta z_{\rm
  wl}=(0.025, 0.097)$ for $z=(1,2)$, respectively.  The weak lensing
uncertainty cannot easily be corrected by more than a factor of $\sim
40\%$ \citep[see][]{dalal03,koc06,gunnarsson06}.  Once the {\it LISA}
uncertainty decreases below this limit, the total redshift error will
not improve.  The redshift distribution of optical galaxies at 27--28
mag is already known from deep observations \citep[e.g][using the
Hubble Ultra Deep Field]{coe06}, with roughly 1--$10\%$ of them
falling in a radial shell of width $\delta z\sim 0.05$ around redshift
$z=1$--2, hence photometric redshifts will offer a factor of $\sim
10$--100 reduction in the number of candidates.

{\em Expected Luminosity Range.}  Since {\it LISA} will also measure
accurate BH masses prior to the merger (the masses are ``fast
parameters'', see \S~\ref{sec:transient}) it may be possible to
roughly estimate the magnitude of the host galaxy from the known
empirical correlations between galaxy properties and SMBH mass.  For
example, extrapolating from the sample of nuclear SMBHs in the local
universe, a $\sim 10^6~{\rm M_\odot}$ SMBH is expected to reside in a
galaxy with a spheroid mass $\sim 10^9~{\rm M_\odot}$, or absolute
magnitude $M_B \sim -16$ \citep{fm00}, corresponding to an apparent
optical magnitude of $m \sim 28$ at $z=1$.

A large fraction of the galaxies will not have luminosities consistent
with the mass--luminosity relationship and the BH masses determined by
LISA, allowing to efficiently reduce the number of counterparts.
Using Eq.~(19) of \citet{ff05}, we find that the $B$-band spheroidal
luminosities are
$M_B=(-12.7^{+1.5}_{-2.2},-15.1^{+1.6}_{-1.1},-17.5^{+1.0}_{-0.65})$
for SMBH masses $M_{\rm SMBH}=2\times 10^{5,6,7}\Msun$. Note, that the
errors are (by far) dominated by the uncertainties in the $M_{\rm
SMBH}$--$M_B$ relationship, and not the LISA measurement errors, with
$\delta M_{\rm SMBH}/M_{\rm SMBH}\lsim 10^{-4}$.  In order to obtain
the total luminosity, we must divide the spheroidal luminosity with
the spheroidal/total luminosity ratio $S/T$.  This ratio depends on
the galaxy morphology, orientation, and luminosity. For the
low--luminosity spheroids that will host the SMBHs with $M_{\rm
SMBH}=10^{5-7}\Msun$, a conservative assumption is $S/T=0.1$--1
\citep{bdfs07}.  Integrating the observed luminosity function of
galaxies between these $M_B$ bounds \citep[using their fits for faint
galaxies at $z=1$, $\Phi^{*}=26.1\times 10^{-4}\Mpc^{-3}$,
$M_B^{*}=-22.4$, $\alpha(z)=-1.12-0.12z$, and $P=-1.00$, $Q=1.72$
given by \cite{Lin99} for $q_0=0.1$, and extrapolating this
relationship to low luminosities]{Ryan07} and using the cosmological
comoving volume element \citep{eis97}
\begin{equation}
 N_{\rm counterpart}(z) = \delta \Omega  \delta z  \frac{\partial^2 V_{\rm co}(z)}{\partial z\partial \Omega}
\int_{M_{B,\min}-2.5 \log(S/T)_{\max}}^{M_{B,\max}-2.5 \log(S/T)_{\min}} \phi(M_B,z) \D M_B
\end{equation}
gives $N_{\rm counterpart}/{\rm deg}^{2}={10^3 \times (7.8,3.7,1.7)}$
and $10^3 \times (10, 3.6,1.2)$ for $M_{\rm SMBH}=2\times
10^{(5,6,7)}\Msun$, $\delta z=0.05$ and for $z=1$ and 2,
respectively. If a bulge/disk decomposition is also available to
constrain $S/T$ \citep{bdfs07}, then $N_{\rm counterpart}$ is reduced
by another factor of $\sim 2$.  Therefore imposing bounds on the range
of both the luminosity and the redshift of the counterpart reduces the
number of counterparts significantly. Incorporating the angular
localization cut above leads to a total of $N_{\rm counterpart}\sim
10^3$--$10^{4}$ at 10 days before merger, $100$--$1000$ at 1 day, and
1--$1000$ at ISCO.

We caution, however, that before applying cuts on candidates, one
should be well aware that the relation between total SMBH mass and
spheroidal luminosity is currently known only for nearby
galaxies, selected by observational feasibility criteria. Clearly,
the relation can be different for rare objects of a transitional
nature, such as SMBHs in the last few weeks of their merger.

{\em Expected velocity dispersion.}  There is observational evidence
for a tight correlation between the SMBH mass, $M$, and the large
scale bulge velocity dispersion of galaxies, $\sigma$ \citep{geb00}.
By contrasting $M$, as measured by LISA, with $\sigma$ obtained from
galactic spectra, a major fraction of the galaxy candidates could be
ruled out.  Using equation~20 in \citet{ff05}, and extrapolating to
lower masses, we find $\sigma=(50\pm 5, 81\pm 4, 130 \pm 1)$, for
$M=2\times 10^{5,6,7} \Msun$, respectively.  In comparison, the
velocity dispersion for observed galaxies ranges between $\sigma\sim
(60$--$400)\,{\rm km/s}$ \citep{sheth03}. Since we expect there will
be many more galaxies with low velocity dispersion, a very efficient
reduction of $N_{\rm counterpart}$ will be only possible for the most
massive detected SMBHs.  Unfortunately, the measurement of velocity
dispersions for faint galaxies is extremely difficult in practice. It
may be feasible only for the larger mass BHs, and presumably after
longer--integration spectra are obtained well after the merger. As
before, we caution that the relation between total SMBH mass and the
velocity dispersion is unknown for rare objects of a transitional
nature, such as SMBHs in the process of coalescence.

{\em Quasar counterparts.}  If the {\it LISA} source is producing a
continuous near--Eddington luminosity with fluctuations at the few
percent level, then most likely, it will appear as a point source
(i.e. a faint quasar) and the host galaxy will not be detectable.  The
bright point source in the nucleus, with $\sim 24$~mag, will outshine
the galaxy.  Kocsis et al. (2006) found that the typical {\it LISA}
error box ($\delta \Omega=0.3\deg^2$, $\delta z=2.5\%$), for a merger
event produced by $\sim 10^6~{\rm M_\odot}$ SMBHs at $z=1$ would
statistically contain $\sim 1$ quasar with at least $10\%$ Eddington
luminosity; the $\sim 30$ times larger area $\sim 2$ weeks before
merger will contain $\sim30$ such quasars.  These quasars could be
identified in advance, and monitored separately for variability with
smaller--field of view instruments.

{\em Orbital Eccentricity as Indication of Circumbinary Gas.}  In 2D
simulations of circumbinary disks, \citet{am02} and
\citet{macfadyen07} find that the interaction of the binary with the
ambient gas produces eccentricity (in the orbit of both the gas and
the binary).  The eccentricity leaves a strong imprint on the GW
waveform \citep{bc04,vm07}.  Moreover, it is a ``fast parameter''
since it modifies the spectrum of the signal on the binary orbital
timescale (see \S~\ref{sec:assumptions}).  Therefore we expect that
the {\it LISA} data--stream should be able to discover such
eccentricity several months before merger and thus alert us to the
occurrence of a gas--driven inspiral.

{\em Spin and Orbital Plane Alignments as Indication of Ambient Gas.}
Prior to the merger, the two SMBHs may reside in two bulges, which are
driven together during the galaxy merger by large--scale torques
\citep[e.g.,][]{dmh96}.  As a result, the orbital plane of the binary
SMBH may be aligned with the plane of the galactic remnant.  Indeed,
in gas rich mergers, the binary plane may align with the plane of the
disk \citep{brm07}. Moreover, these authors argue that the interaction
with the gas also aligns the individual BH spins in the direction of
the orbital angular momentum. Therefore the relative orientation of
the binary orbit, spin vectors, and embedding galaxy may be valuable
observables. In particular, if the measured spin vectors are found to
align with the orbital angular momentum, it may indicate that a
sufficient amount of gas was present in the history of the
merger. Moreover, in the case of alignment, the direction of the
recoil kick at merger is in the orbital plane, making the
kick--triggered variability discussed in \S~\ref{sec:transient} more
plausible.  Conversely, misaligned spins may indicate that the merger
is dry, and an EM signal might not as easily accompany the GW
event. Note that a $\pm 20\deg$ triple alignment between the spins and
orbital momentum by chance has a mere $9\times 10^{-4}$ probability.
In case there are several coincident inspirals present in the LISA
data-stream, a favorable EM follow-up candidate could thus be chosen
according to a criterion for potential EM activity based on the
observed relative alignments.

\begin{figure*}[tbh]
\centering
\mbox{\includegraphics[height=4.85cm]{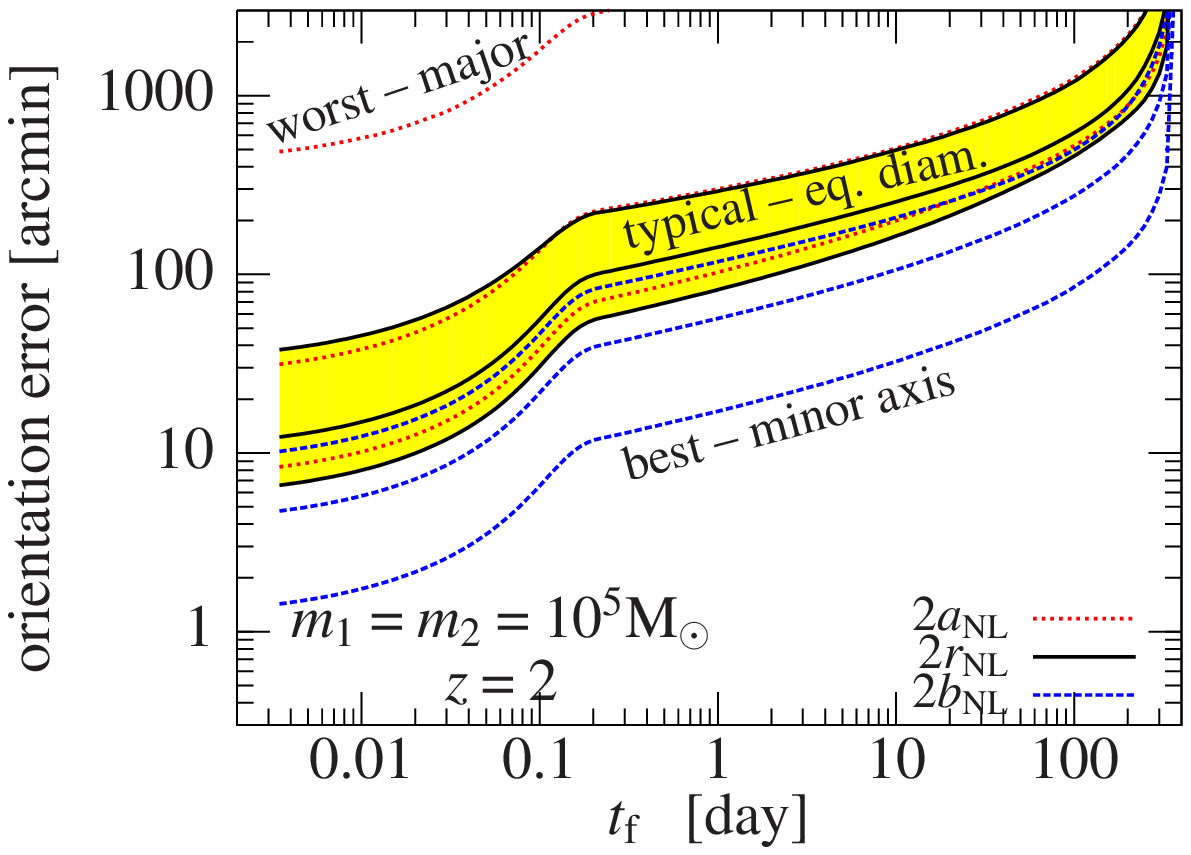}}
\mbox{\includegraphics[height=4.85cm]{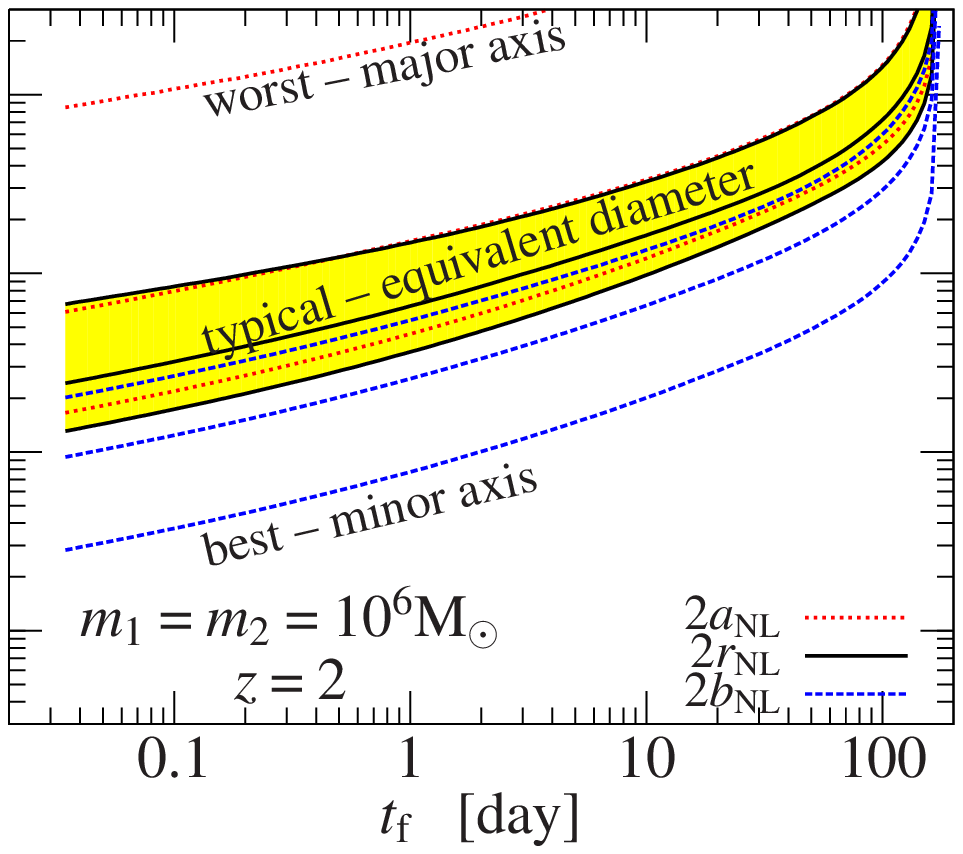}}
\mbox{\includegraphics[height=4.85cm]{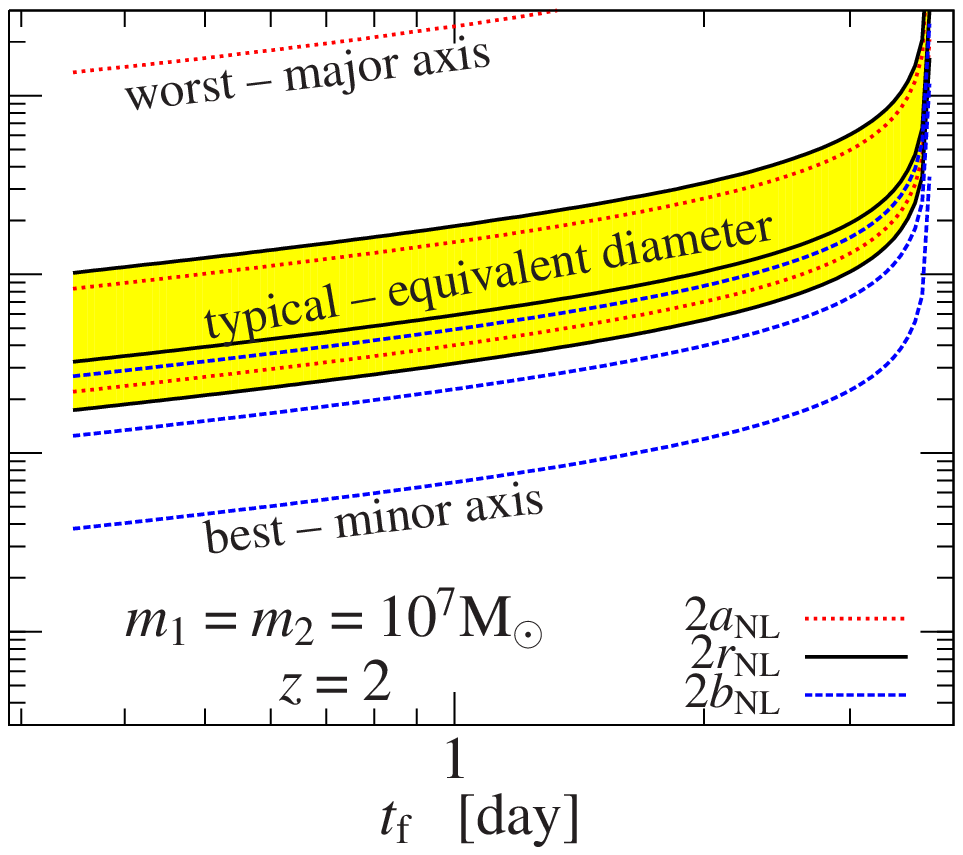}}\\
\mbox{\includegraphics[height=4.85cm]{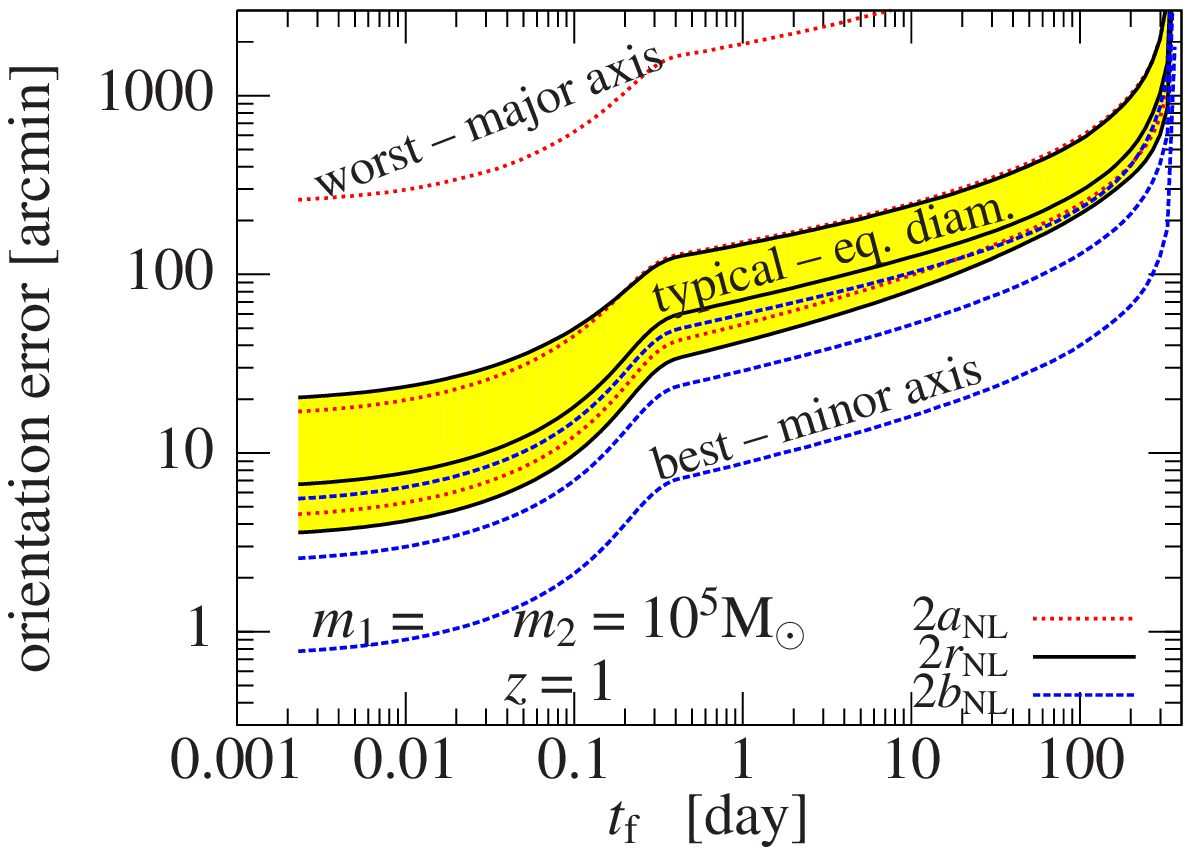}}
\mbox{\includegraphics[height=4.85cm]{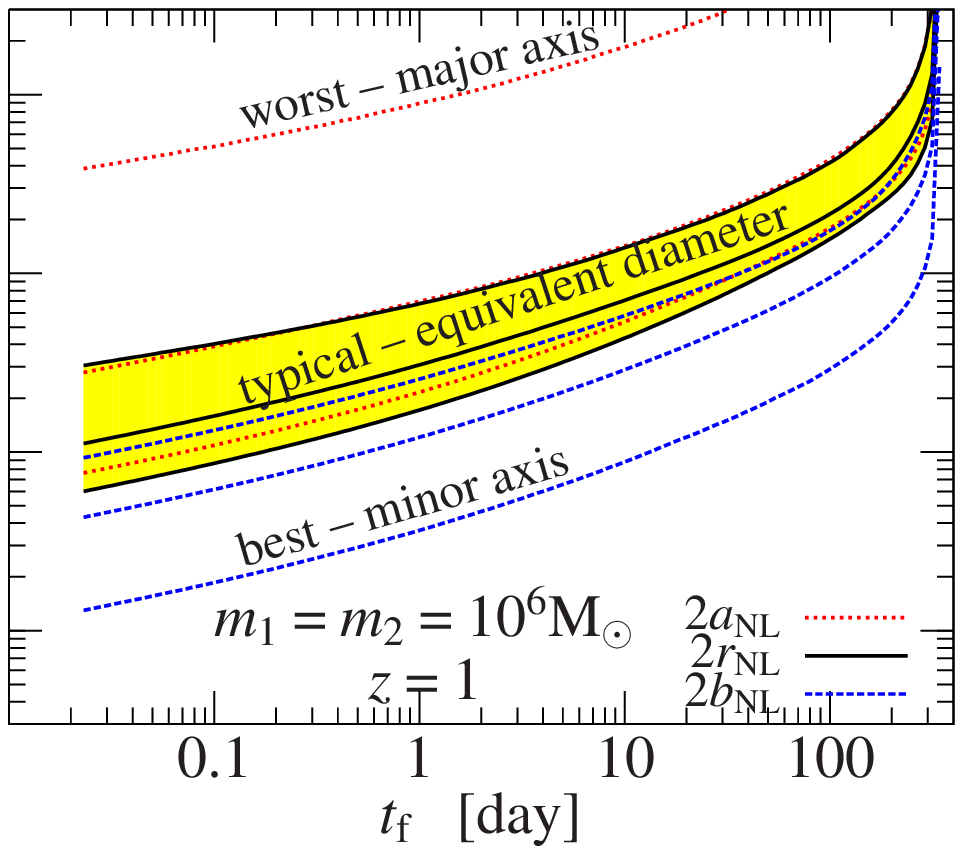}}
\mbox{\includegraphics[height=4.85cm]{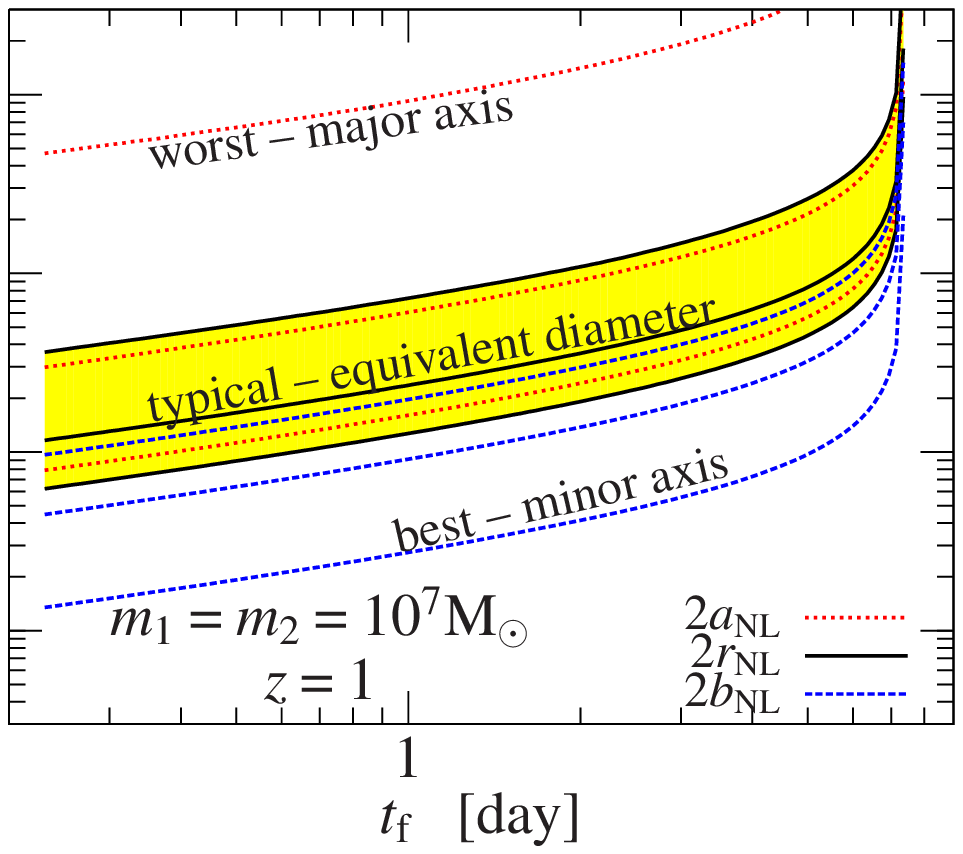}}
\caption{\label{f:orbit} Evolution with pre-ISCO look--back time,
  $\tf$, of LISA errors for the orientation of the orbital plane of
  the SMBH binary relative to the line-of-sight. The inclination and
  rotation uncertainties are correlated leading to an error ellipsoid
  with major axis, $2a_{\rm NL}$, equivalent diameter $2r_{\rm NL}$ and
  minor axis $2b_{NL}$.  Best, typical, and worst cases for random
  orientation events represent the $10\%$, $50\%$, and $90\%$ levels
  of cumulative error distributions, respectively. The range spanned
  by the $2r_{\rm NL}$ distribution is highlighted for clarity.  The
  figure shows that the binary inclination angle can be precisely
  localized very early on, to within degrees, allowing a selection of
  prime host galaxy candidates that have a similar inclination.  }
\end{figure*}

In Figure~\ref{f:orbit}, we show the accuracy with which we will know
the inclination of the SMBH orbital angular momentum, in advance of
the merger, as specified by the two angles $(\theta_L,\phi_L)$. When
considering the individual EM candidates in the LISA error volume,
one-by-one, the EM information allows to set the sky position errors
$(\delta\theta_N,\delta\phi_N)$ to essentially negligible
values. Therefore, it is meaningful to calculate the 2D error
ellipsoid for $(\theta_L,\phi_L)$ by neglecting correlations with
$\delta\theta_N$ and $\delta\phi_N$.  Figure~\ref{f:orbit} then shows
that we will know the inclination remarkably well in advance. One may
therefore envision focusing on galaxy candidates that exhibit
inclinations similar to that of the SMBH binary. This will be feasible
only if the host galaxy is actually detected and resolved (i.e. if the
SMBH binary does not produce near--Eddington emission whose PSF
completely hides the galaxy).  The angular size of small elliptical
galaxies at $z=1-3$ with optical magnitudes $\sim 27$ is expected to
be 0.1--0.5 arcseconds \citep[see, e.g., simulations by ][]{kw03},
which cannot be resolved from the ground. However, by $z\sim1 $, most
of the host spheroids may be surrounded by much larger disks.  For
example, the 20~kpc diameter disk of a galaxy like our own Milky Way
would extend beyond $2$ arcseconds at $z=1-3$, and could thus be
resolved.  Again, one can envision separately monitoring with smaller
field--of--view instruments a few of those galaxies (if any) whose
disks appear to be closely aligned with the orbital plane of the SMBH
binary.

In conclusion, the alignment of spins (indicative of the presence of
gas) and the alignment of orbital momentum with the plane of the
galaxy can be very helpful in choosing the most likely host galaxy
candidates.  Note, that a $\pm 20\deg$ coincidence between the GW and
EM measured orbital orientation has a $3\%$ probability, thereby
allowing a very effective cut to be made on the candidates.

{\em Galaxy Overdensity as Indication of Enhanced Merger Rate.}
Another possibility is to focus on galaxy clusters within the few
square degrees of interest.  There may be a few to a few tens of known
galaxy clusters in this area. Combining with the photometric redshift
information, $\Delta z=0.05$, there are on average $N_{\rm
  cluster}/\deg^{2}=0.2$ and 0.01 clusters within this shell, at $z=1$
and 2, respectively \citep[e.g.][]{hhm01}. The probability that a
randomly chosen galaxy has undergone a recent merger is significantly
larger in a cluster than in the field; again, it would make sense to
separately monitor the dense cluster environments with smaller FOV
telescopes.

\section{Motivations and New Fundamental Tests of Gravity}
\label{sec:graviton}

We have argued that real-time triggered searches for EM counterparts
to SMBH binary merger events observed by LISA will be technically
feasible, at least for a subset of all such events. Observational
strategies to achieve such identifications constitute major
astronomical endeavors, however.  It is thus important to clarify the
nature of the additional science that could be enabled by real-time
identifications.  We describe two important motivations for pre-merger
localizations here.

The first motivation is simply coincidence. Post-merger identification
of host galaxies will likely rely on active nucleus variability or on
a statistical association with rare galactic objects (e.g. quasars,
ULIRGs, X-ray or starburst galaxies; \citealt{koc06,dot06}). Depending
on the nature of the AGN variability or the specific galactic
attribute one is willing to associate with a SMBH binary merger,
possibilities for chance coincidences may remain uncomfortably high.
The combination of space (sky location) and time coincidence for a
prompt EM counterpart that is gravitationally-timed with a SMBH merger
should considerably reduce the risks of chance coincidences and would
thus likely be the most secure way of identifying a truly unique
counterpart and host galaxy. This is particularly important in this
context, with observations of a rare transient astrophysical
phenomenon that has never been observed before, whether
gravitationally or electromagnetically. Secure counterpart
identifications are also crucial if one is to perform new fundamental
tests of gravitational physics with these events, such as comparisons
between electromagnetic and gravitational Hubble diagrams
\citep{dm07}.

The second important motivation is that the detection of prompt
electromagnetic counterparts itself may provide additional
opportunities for testing gravitational physics on cosmological
scales. We highlight here one such new test, based on the measurement
of photon and graviton arrival times from the same cosmological
source. As we already emphasized (see \S\ref{sec:transient}) the most
violent and energetic phase of the SMBH binary merger will be the
final coalescence of the two BHs. Let us assume for simplicity that a
luminous burst of light is generated at the same time as the most
luminous burst of GWs is emitted by the coalescing binary system. If
gravitons propagate at a speed lower than that of light, a delay in
the arrival time of the burst of gravitons, relative to the burst of
photons, may become apparent after propagation over cosmological
distances.

One specific scenario in which this is expected is if the graviton
possesses a non-zero rest-mass \citep[e.g,
][]{pf39}.\footnote{Building a satisfactory Lorentz-invariant theory
  of massive gravity remains a challenge. Possibly more successful
  Lorentz-violating versions have also been proposed \citep[e.g,
  ][]{dub05}.}  Assuming that Lorentz invariance is satisfied by the
massive gravity theory under consideration, we can infer the bound
that can be set on the graviton mass, for a given value of the delay
between photon and graviton arrival times. Let us write the light
travel time to a cosmological standard siren as $\eta$ \citep[i.e. the
conformal time, e.g. ][]{eis97} and the velocity of gravitons as $c_g
< c,$ where $c$ is the speed of light. By definition, we have
\begin{equation} \label{eq:delay}
\frac{c_g}{c} = 1- \frac{\Delta t} {\eta},
\end{equation}
where $\Delta t > 0$ is the delay in graviton arrival times. The
corresponding Lorentz factor is
\begin{equation} \label{eq:gamma}
\gamma \equiv \frac{1}{\sqrt{1-(c_g/c)^2}} \simeq \sqrt{\frac{\eta}{2
\Delta t}},
\end{equation}
so that from $\gamma m_0 c^2 = h f $, one could deduce the graviton
rest mass $m_0$ from the GW frequency, $f$, and any measured delay,
$\Delta t$. Note that, using the conformal time $\eta$ can be
expressed with redshift for a given cosmology (e.g. \citealt{eis97}),
and no other cosmological factors enter this calculation.

Without any special treatment, it appears unlikely the burst of GWs
can be isolated and thus accurately timed to better than about a
dynamical timescale for the coalescing SMBH binary (i.e., one final GW
cycle). This timing uncertainty will thus set a limit on the minimum
graviton rest mass that can be constrained from delays in arrival
times. Adopting the inverse of the frequency at ISCO as the limiting
timing uncertainty, $f_{\rm isco} = 2.2\times 10^{-(2,3,4)}\Hz$, for
an SMBH binary with total mass $M=10^{5,6,7}\Msun$ at $z=1$,
respectively, for which $\eta = 11.6$~Gyr, we find that bounds on the
graviton mass based on arrival time delays are possible as long as
$m_0 \gsim (14, 4.5, 1.4) \times 10^{-25}$~eV or in terms of Compton
wavelength $\lambda_{\rm c}\lsim (28,89,280)$~pc, respectively.  This
bound on the graviton mass is less stringent, than the $1~$kpc value
derived from the deviations that would emerge in the phasing of
massive GW waveforms \citep{bbw05,will04}. In addition, any systematic
delay in the emission of the burst of photons, which must causally
follow the gravitational event and thus the burst of gravitons, could
significantly reduce the quality of the constraints on $m_0$ based on
differences in arrival times between photons and gravitons. In fact,
in the absence of any quantitative information on such systematic
delays at the source, one will be inevitably reduced to
model--dependent statements, with specific emission scenarios, unless
observations were to reveal an unambiguous propagation effect (with
the burst of photons significantly preceding the burst of gravitons).

In principle, however, a possibility exists for more accurate relative
timing of the photon and graviton signals. Before coalescence, gas
present in the near environment of the SMBH binary would be
gravitationally perturbed in such a way that it could radiate a
variable signal with a period closely matching that of the
leading-order quadrupolar perturbation induced by the coalescing
binary. In this case, independently of the details of the
electromagnetic emission process, it may be possible to match the
periods of the electromagnetic and GW signals. The
offset in phase between the Fourier components of the two signals
with similar frequencies could thus be used to effectively calibrate
the intrinsic delay in electromagnetic emission at the source. Any
drift in arrival-time with frequency between the gravitational and
electromagnetic chirping signals, as the source spans about a decade
in GW frequency during the last 2 weeks before merger (see
Fig.~\ref{f:torbit}), could then be attributed to a fundamental
difference in the way photons and gravitons propagate on cosmological
distances. Another immediate advantage of this frequency matching is
that there is no need to isolate a luminous burst of GWs and to
associate it with a corresponding burst of photons in the observed
signals. Instead, the various phases of late inspiral and coalescence
can be tracked via the GW signal, so that the relative timing of the
gravitational and electromagnetic signals may be known to within a
fraction of the binary's dynamical time. Of course, it is difficult to
assess the accuracy of this matching and tracking with frequency
without a more detailed analysis. However, what is clear from our
previous discussion is that if the delay in electromagnetic emission
at the source can be efficiently calibrated and the absolute timing of
both signals can be reduced to less than the binary's orbital period,
then the direct comparison of signals could provide constraints on the
graviton mass, $m_0$, of comparable or even better quality than those
obtained from the phasing of the GWs alone \citep{bbw05,will04}.

It is important to emphasize that the above limits on the value of the
graviton mass, $m_0$, are strongly theory-dependent. In our
discussion, as in the analyses of \citet{bbw05} and \citet{will04},
it was assumed that the finite value of the graviton mass has
negligible effects on the rate of decay of the binary or other
observable properties of the GW signal (such as
polarization). Furthermore, Lorentz invariance was explicitly used to
relate the delay in arrival times of gravitons to their rest mass,
$m_0$. These various assumptions can be violated in various theories
of modified gravity, whether massive or not. We will mention a few
possibilities of interest here in that they could be
phenomenologically tested.

In braneworld models with extra dimensions available to gravity beyond
a large, cosmological ``cross-over'' scale \citep[e.g.,][]{DGP}, from
the Kaluza-Klein mechanism, massless higher-dimensional gravitons may
start behaving like 4D-massive particles once they travel distances
comparable to the theory cross-over scale. There would not be any
single rest mass associated with 4D-massive gravitons, so that the
specific constraints discussed above or in \citet{bbw05} would not
strictly apply. That is, after cosmological propagation, the relation
between arrival time and GW frequency may be more complex than assumed
in equation~(\ref{eq:gamma}). Even though Lorentz invariance has been
extensively tested for standard model fields, the possibility remains
that Lorentz symmetry is violated in the gravity sector, especially on
cosmological scales.  Such apparent gravity sector Lorentz violations
emerge in braneworld scenarios, if the phenomenology is incorrectly
interpreted from a 4D-spacetime point of view
\citep[e.g.][]{ckr02,ceg01}.  Independently of higher-dimensional
scenarios, specific Lorentz-violating theories of gravity in 4D have
also been proposed in recent years, e.g. with a vector-tensor
character \citep[e.g.][]{bek04,jm04}.

In this context, it is thus significant that electromagnetic
counterparts to SMBH binary mergers, together with the opportunity to
match and track the gravitational and electromagnetic signals in
frequency, may offer unique tests of Lorentz violations in the gravity
sector. It may be possible, as the SMBH binary decays toward final
coalescence, spanning a range of frequencies, to measure the delays in
graviton vs photon arrival times as a function of increasing GW
frequency, $f$. Therefore, the consistency of Eq.~(\ref{eq:delay})
with the special relativistic relation $\gamma m_0 c^2 = h f $ could
be tested explicitly, potentially revealing empirical violations of
Lorentz invariance for gravitons propagated over cosmological
scales.
Clearly, to have any chance to perform such new tests of gravitational
physics, one will need to identify the electromagnetic counterparts as
early as possible. This may be one of the strongest motivations behind
ambitious efforts to localize these rare, transient events well before
final coalescence.

\section{Discussion and Implications}
\label{sec:discuss}

In this paper, we have considered the possibility of localizing
coalescing SMBH binaries in real time, by using the gravitational
waves detected by {\it LISA} to trigger searches for electromagnetic
counterparts using wide--field instruments.  The idea presented here
is that such searches may reveal a time-variable source -- showing
either quasi--periodic brightness variations during the inspiral
stage, or a transient ``burst'' during the coalescence itself -- and
that this may allow identification of a unique counterpart.

We first analyzed the behavior of the time--evolving localization
error prior to the merger. As the binary evolves in time, the
signal-to-noise ratio, $S/N$, per unit time of the GW detection
increases rapidly. Therefore the measurement precision of physical
parameters, such as the 3D position, inclination, and spins can
improve quickly as the merger approaches.  We have calculated the size
and shape of the 3D localization ellipsoid, and its distribution for
arbitrary binary orientations and sky positions, as well as how these
uncertainty distributions evolve before the merger.  We found that for
typical sources at $z=1$--2, $\sim 10$ days before merger, the
distance measurement precision reaches its limiting value $\delta
z_{wl}\sim (2$--$10)\%$ imposed by weak gravitational lensing.
Redshift errors subsequently decouple from the sky position errors,
and the 3D localization ellipsoid aligns with the line of sight. The
sky position error decreases below $10\deg^2$, 10 days before merger,
for a large range of masses, $10^5\leq M/\Msun\leq 5\times 10^6$, at
redshifts $z\leq 1.5$.  We have also found that the sky position error
ellipse is nearly circular after the first few weeks of observation,
except for the final $\sim$day before the merger, especially for the
small ellipsoids with favorable sky positions and inclinations.
During the final day, the minor axis of the sky ellipse shrinks more
quickly, especially for large BH spins \citep{lh07}, creating an
eccentric sky position error ellipse.  Finally, we found that the
source position can be important: localization improves quickly for
sources that lie in the plane defined by the {\it LISA} triangle, but
remains much poorer in the perpendicular direction.

We then analyzed the prospects for electromagnetic monitoring of this
sky area to look for prompt counterparts and host galaxy
candidates. It will be possible to estimate the time of the merger to
better than an hour precision already several months in advance of
merger.  The expected number of galaxies that would be present in the
GW localization region is up to $10^6$ during the last weeks.
However, we have shown that using photometric redshift information and
luminosity cuts consistent with the empirical black hole mass --
luminosity measurements, one can decrease the number of galaxy
counterpart candidates considerably, down to a number $\sim
10^3$--$10^4$ approximately 10 days before merger, $\sim 100$--$1000$
at 1 day, and $\sim 1$--$1000$ at ISCO. These galaxies will be
typically very faint, with absolute magnitudes between $M_B=(-18, -16,
-14)\pm 2$ for $M=10^{5,6,7}\Msun$. Moreover, the GW localization
volume will enclose $\sim 30$ quasar counterpart candidates at 10 days
before merger, decreasing to $\sim 1$ quasar close to ISCO.  We find
that both the orbital inclination and the spins can be measured with
GWs to an excellent precision for each counterpart candidate in the
field, several months before the merger.  Using this information, it
may be possible to assess, in advance, which SMBH binaries will
receive a recoil with a direction (and known magnitude) that is likely
to plunge them into the surrounding gaseous disk, favoring a transient
EM signal. If SMBH binary orbits align with the plane of the galactic
disk, then measuring the disk orientation to $\pm20\deg$ would have a
potential for further reducing the number of counterpart candidates by
typically more than an order of magnitude.

The best hope to find a merger event EM counterpart may be to look for
variable emission produced by gas near the BHs. EM variability can be
expected if there is ambient gas surrounding the coalescing binary, on
the timescale of the periodic gravitational perturbation, $\sim
0.1$--$10$~hr. An electromagnetic transient might take place shortly
after the merger due to GW recoil or the sudden central rest-mass
loss.  The resulting counterpart candidates should be monitored for
any type of EM transients or periodic variability whose period
and time dependence is correlated with that of the known
gravitational perturbation.  The basic requirements of a wide field
of view, and fast detectors, are similar to those of searches being
planned for distant cosmological supernovae.  For typical LISA
sources, a triggered EM counterpart search campaign will require
monitoring a several--square degree area, and could aim for
variability at the 24--27 mag level in optical bands,
corresponding to 1-10\% of the Eddington luminosity for prime {\it
LISA} sources with $\sim (10^6$--$10^7)\,{\rm M_\odot}$ BHs at
$z=1-2$, on time--scales of minutes to hours.  A cross--correlation
of the phase of any variable EM signal with the quasi--periodic
gravitational waveform over 10-1000 cycles may aid the detection.  The
optical emission from near the BHs may be obscured by large columns of
gas and dust during the merger, in which case monitoring a similar
area, to similar flux levels in X--ray, infrared or radio bands may be
required.

The real--time monitoring envisioned here could well make the
difference between detecting and missing the prompt EM counterpart.
Our results therefore also constitute a strong scientific case
for allowing data from LISA to be optionally downloaded on $\sim$
hourly timescales, to permit near--real--time analysis when a
particularly interesting event is noticed.

The secure identification of the EM counterpart to a GW event
would indeed have far--reaching implications.  The identification of
a counterpart would provide observational insights into poorly
understood processes related to SMBH binary accretion and transients
during a general relativistic merger involving the presence of gas.
Another tantalizing application of such observations is the
possibility to constrain, in new ways, models of gravity on
cosmological scales and the geometry of the universe with these
events.  Comparing arrival times between the EM and GW signals may
allow stringent bounds to be placed on the mass of the graviton and
could enable empirical tests of Lorentz violations in the gravity
sector.

\acknowledgments We thank Alessandra Buonanno, Scott Hughes, Frans
Pretorius, Abraham Loeb, and Andr\'as P\'al for helpful discussions,
and Zsolt Frei, Ryan Lang, Daniel Chung and George Djorgovski for useful comments on the
manuscript.  BK acknowledges support from the ITC at Harvard
University, the \"Oveges J\'ozsef Fellowship, and OTKA grant
no. 68228.  KM was supported in part by the National Science
Foundation under Grant No. PHY05-51164 (at KITP). ZH acknowledges
partial support by NASA through grant NNG04GI88G, by the NSF through
grant AST 05-07161, and by the Pol\'anyi Program of the Hungarian
National Office for Research and Technology (NKTH).

\end{document}